\RequirePackage{fix-cm}
\documentclass[twocolumn]{svjour3}          
\smartqed  
\usepackage{array}
\usepackage{xcolor}
\usepackage{subcaption}
\usepackage{amsmath}
\usepackage{amssymb}
\usepackage{graphicx}
\usepackage{xspace}
\usepackage{mathtools}
\usepackage{multirow}
\usepackage{booktabs}

\usepackage{mathptmx}      
\usepackage{latexsym}
\newcommand{\reffigure}[1]{Figure~\ref{#1}}
\newcommand{\refsection}[1]{Section~\ref{#1}}

\newcommand{\reftable}[1]{Table~\ref{#1}}

\newcommand{\refappendix}[1]{Appendix}

\newcommand{\btree}{B$^+$-tree\xspace}
\newcommand{\btrees}{B$^+$-trees\xspace}

\newcommand{\primary}{$\clubsuit$\xspace}
\newcommand{\secondary}{$\vartriangle$\xspace}
\newcommand{\zap}[1]{}
\DeclarePairedDelimiter{\ceil}{\lceil}{\rceil}

\newcolumntype{P}[1]{>{\centering\arraybackslash}p{#1}}

\begin{document}
\title{LSM-based Storage Techniques: A Survey
}

\author{Chen Luo         \and
        Michael J. Carey
}


\institute{Chen Luo \at
			University of California, Irvine\\
            \email{cluo8@uci.edu}           
            \and
			Michael J. Carey \at
			University of California, Irvine\\
			\email{mjcarey@ics.uci.edu}
}

\date{Received: date / Accepted: date}

\maketitle

\begin{abstract}
Recently, the Log-Structured Merge-tree (LSM-tree) has been widely adopted
for use in the storage layer of modern NoSQL systems.
Because of this, there have been a large number of research efforts, from both the database community
and the operating systems community,
that try to improve various aspects of LSM-trees.
In this paper, we provide a survey of recent research efforts on LSM-trees so that readers can
learn the state-of-the-art in LSM-based storage techniques.
We provide a general taxonomy to classify the literature of LSM-trees,
survey the efforts in detail, and discuss their strengths and trade-offs.
We further survey several representative LSM-based open-source NoSQL systems
and discuss some potential future research directions resulting from the survey.
 
\end{abstract}

\section{Introduction}
The Log-Structured Merge-tree (LSM-tree) has been widely adopted in the storage layers of modern NoSQL systems,
including BigTable~\cite{bigtable}, Dynamo~\cite{dynamo2007},
HBase~\cite{hbase}, Cassandra~\cite{cassandra}, LevelDB~\cite{leveldb},
RocksDB~\cite{rocksdb}, and AsterixDB~\cite{asterixdb-storage2014}.
Different from traditional index structures that apply in-place updates,
the LSM-tree first buffers all writes in memory and subsequently flushes them to disk and merges them using sequential I/Os.
This design brings a number of advantages,
including superior write performance, high space utilization, tunability, and simplification of concurrency control and recovery.
These advantages have enabled LSM-trees to serve a large variety of workloads.
As reported by Facebook~\cite{rocksdb-space2017}, RocksDB, an LSM-based key-value store engine,
has been used for real-time data processing~\cite{facebook-realtime2016}, graph processing~\cite{dragon}, stream processing~\cite{facebook-realtime2016},
and OLTP workloads~\cite{myrocks}.

Due to the popularity of LSM-trees among modern data stores,
a large number of improvements on LSM-trees have been proposed by the research community;
these have come from both the database and operating systems communities.
In this paper, we survey these recent research efforts on improving LSM-trees,
ranging from key-value store settings with a single LSM-tree to more general database settings with secondary indexes.
This paper aims to serve as a guide to the state-of-the-art in LSM-based storage techniques for researchers, practitioners, and users.
We first provide a general taxonomy to classify the existing LSM-tree improvements based on the specific aspects that they attempt to optimize.
We then present the various improvements in detail and discuss their strengths and trade-offs.
To reflect how LSM-trees are being used in real systems, we further survey five representative LSM-based open-source NoSQL systems, including LevelDB~\cite{leveldb}, RocksDB~\cite{rocksdb}, HBase~\cite{hbase}, Cassandra~\cite{cassandra}, and AsterixDB~\cite{asterixdb2014}.
Finally, we also identify several interesting future research directions
as the result of categorizing the existing LSM-tree improvements.

The reminder of this paper is organized as follows.
\refsection{sec:lsm-basics} briefly reviews the history of LSM-trees and presents the basics of today's LSM-tree implementations.
\refsection{sec:lsm-improvements} presents a taxonomy of the proposed LSM-tree improvements and surveys the existing work using that taxonomy.
\refsection{sec:lsm-systems} surveys some representative LSM-based NoSQL systems, focusing on their storage layers.
\refsection{sec:future-direction} reflects on the result of this survey, identifying several outages and opportunities for future work on LSM-based storage systems.
Finally, \refsection{sec:conclusion} concludes the paper.

\section{LSM-tree Basics}
\label{sec:lsm-basics}
In this section, we present the background of LSM-trees.
We first briefly review of the history of work on LSM-trees.
We then discuss in more detail the basic structure of LSM-trees as used in today's storage systems.
We conclude this section by presenting a cost analysis of writes, reads, and space utilization of LSM-trees.

\subsection{History of LSM-trees}
\begin{figure}
	\centering
	\includegraphics[width=\linewidth]{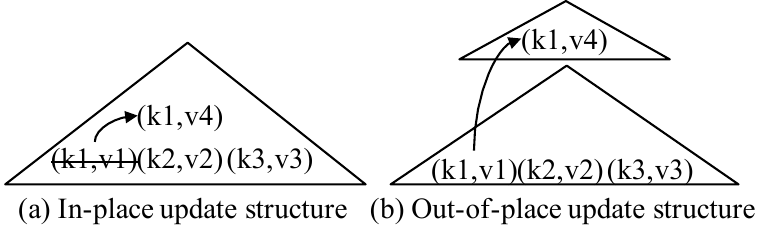}
	\vspace{-0.2in}
	\caption{Examples of in-place and out-of-place update structures: each entry contains a key (denoted as ``k'') and a value (denoted as ``v'')}
	\label{fig:in-out-place}
\end{figure}

In general, an index structure can choose one of
two strategies to handle updates, that is, in-place updates and out-of-place updates.
An in-place update structure, such as a \btree, directly overwrites old records to store new updates.
For example in \reffigure{fig:in-out-place}a, to update the value associated with key k1 from v1 to v4,
the index entry (k1, v1) is directly modified to apply this update.
These structures are often read-optimized since only the most recent version of each record is stored.
However, this design sacrifices write performance, as updates incur random I/Os.
Moreover, index pages can be fragmented by updates and deletes, thus reducing the space utilization.

In contrast, an out-of-place update structure, such as an LSM-tree, always stores updates into new locations
instead of overwriting old entries.
For example in \reffigure{fig:in-out-place}b, the update (k1, v4) is stored into a new place instead of updating the old entry (k1, v1) directly.
This design improves write performance since it can exploit sequential I/Os to handle writes.
It can also simplify the recovery process by not overwriting old data.
However, the major problem of this design is that read performance is sacrificed 
since a record may be stored in any of multiple locations.
Furthermore, these structures generally require a separate data reorganization process to improve storage and query efficiency continuously.

The idea of sequential, out-of-place updates is not new; it has been successfully applied to database systems since the 1970s.
Differential files~\cite{diff-files1976}, presented in 1976, were an early example of an out-of-place update structure.
In this design, all updates are first applied to a \emph{differential file},
which is merged with the \emph{main file} periodically.
Later, in the 1980s, the Postgres project~\cite{postgres1987} pioneered the idea of log-structured database storage.
Postgres appended all writes into a sequential log, enabling fast recovery and ``time-travel'' queries.
It used a background process called the \emph{vacuum cleaner} to continuously garbage-collect obsolete records from the log.
Similar ideas have been adopted by the file system community to fully utilize disk write bandwidth,
such as in the Log-Structured File System (LFS)~\cite{lfs1992}.

\begin{figure}[b]
	\centering
	\includegraphics[width=0.9\linewidth]{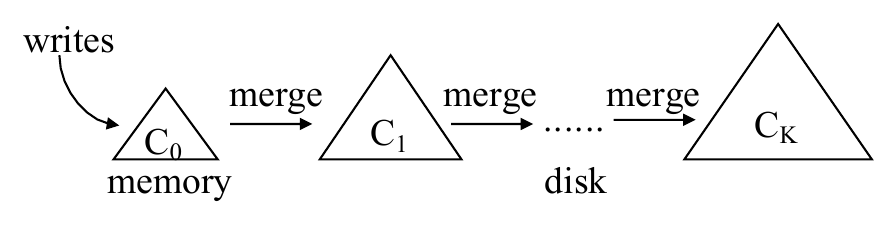}
	\caption{Original LSM-tree design}
	\label{fig:original-lsm}
\end{figure}

Prior to the LSM-tree, the approaches to log-structured storage suffered from several key problems.
First and foremost, storing data into append-only logs leads to low query performance,
as related records are scattered across the log.
Another problem is low space utilization due to obsolete records that have not yet been removed.
Even though various data reorganization processes were designed, there was no principled cost model
to analyze the trade-offs among the write cost, read cost, and space utilization,
which made early log-structured storage hard to tune;
data reorganization could easily become a performance bottleneck~\cite{lfs-phd1992}.

The LSM-tree~\cite{lsm1996}, proposed in 1996, addressed these problems by designing a merge process
which is integrated into the structure itself,
providing high write performance with bounded query performance and space utilization.
The original LSM-tree design contains a sequence of components $C_0$, $C_1$, $\cdots$, $C_k$, as shown in \reffigure{fig:original-lsm}.
Each component is structured as a \btree.
$C_0$ resides in memory and serves incoming writes, while all remaining components $C_1$, $\cdots$, $C_k$ reside on disk.
When $C_i$ is full, a rolling merge process is triggered to merge a range of leaf pages from $C_i$ into $C_{i+1}$.
This design is often referred to as the leveling merge policy~\cite{dostoevsky2018,monkey2017} today.
However, as we shall see later, the originally proposed rolling merge process is not used by today's LSM-based storage systems
due to its implementation complexity.
The original paper on LSM-trees~\cite{lsm1996} further showed that under a stable workload, where the number of levels remains static,
write performance is optimized when the size ratios $T_i = |C_{i+1}|/|C_i|$ between all adjacent components are the same.
This principle has impacted all subsequent implementations and improvements of LSM-trees.

In parallel to the LSM-tree, Jagadish et al.~\cite{step-merge1997} proposed a similar structure with the stepped-merge policy
to achieve better write performance.
It organizes the components into levels, and when level $L$ is full with $T$ components,
these $T$ components are merged together into a new component at level $L+1$.
This policy become the tiering merge policy~\cite{dostoevsky2018,monkey2017} used in today's LSM-tree implementations.

\subsection{Today's LSM-trees}
\label{sec:lsm-basics-today}

\subsubsection{Basic Structure}
\label{sec:lsm-basic-structure}
Today's LSM-tree implementations still apply updates out-of-place to reduce random I/Os.
All incoming writes are appended into a memory component.
An insert or update operation simply adds a new entry,
while a delete operation adds an anti-matter entry indicating that a key has been deleted.
However, today's LSM-tree implementations commonly exploit the immutability of disk components\footnote{Also referred to as \emph{runs} in the literature.} to simplify concurrency control and recovery.
Multiple disk components are merged\footnote{Also referred to as \emph{compaction} in the literature.}
together into a new one without modifying existing components.
This is different from the rolling merge process proposed by the original LSM-tree~\cite{lsm1996}.

Internally, an LSM-tree component can be implemented using any index structure.
Today's LSM-tree implementations typically organize their memory components using a concurrent data structure
such as a skip-list or a \btree, while they organize their disk components using \btrees or sorted-string tables (SSTables).
An SSTable contains a list of data blocks and an index block;
a data block stores key-value pairs ordered by keys, and the index block stores the key ranges of all data blocks.

A query over an LSM-tree has to search multiple components to perform reconciliation, that is, to find the latest version of each key.
A \emph{point lookup query}, which fetches the value for a specific key, can simply search all components one by one,
from newest to oldest, and stop immediately after the first match is found.
A \emph{range query} can search all components at the same time, feeding the search results into a priority queue to perform reconciliation.

\begin{figure}
	\centering
	\begin{subfigure}{0.425\textwidth}
			\includegraphics[width=\linewidth]{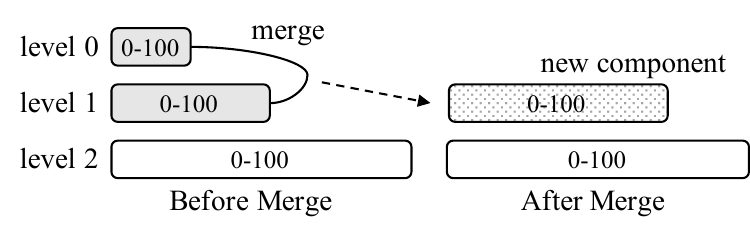}
			\vspace{-0.2in}
			\caption{Leveling Merge Policy: one component per level}
			\label{fig:leveling}
	\end{subfigure}
	\hfill
	\begin{subfigure}{0.425\textwidth}
		\includegraphics[width=\linewidth]{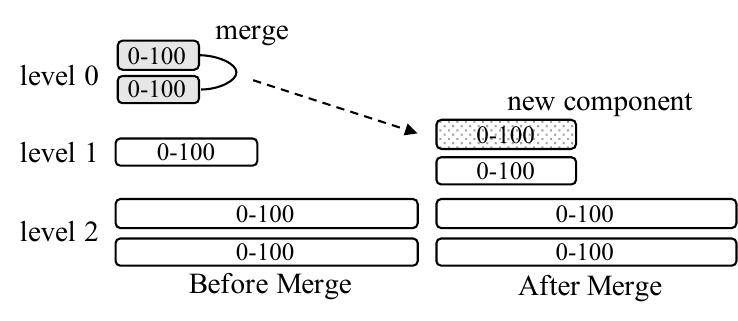}
		\vspace{-0.2in}
		\caption{Tiering Merge Policy: up to T components per level}
		\label{fig:tiering}
	\end{subfigure}
	\caption{LSM-tree merge policies}
	\label{fig:lsm-merge-policy}
\end{figure}

As disk components accumulate over time, the query performance of an LSM-tree tends to degrade since more components must be examined.
To address this, disk components are gradually merged to reduce the total number of components.
Two types of merge policies are typically used in practice~\cite{dostoevsky2018,monkey2017}.
As shown in \reffigure{fig:lsm-merge-policy}, both policies organize disk components into logical levels (or tiers)
and are controlled by a size ratio $T$.
Each component is labeled with its potential key range in the figure.
In the leveling merge policy (\reffigure{fig:leveling}), each level only maintains one component,
but the component at level $L$ is $T$ times larger than the component at level $L-1$.
As a result, the component at level $L$ will be merged multiple times with incoming components at level $L-1$
until it fills up, and it will then be merged into level $L+1$.
For example in the figure, the component at level 0 is merged with the component at level 1, which will result in a bigger component at level 1.
In contrast, the tiering merge policy (\reffigure{fig:tiering}) maintains up to $T$ components per level.
When level $L$ is full, its $T$ components are merged together into a new component at level $L+1$.
In the figure, the two components at level 0 are merged together to form a new component at level 1.
It should be noted that if level $L$ is already the configured maximum level, then the resulting component remains at level $L$.
In practice, for a stable workload where the volume of inserts equal the volume of deletes,
the total number of levels remains static\footnote{Even for an append-mostly workload, the total number of levels will grow extremely slowly since the maximum number of entries that an LSM-tree can store increases exponentially with a factor of $T$ as the number of levels increases.}.
In general, the leveling merge policy optimizes for query performance since there are fewer components to search in the LSM-tree.
The tiering merge policy is more write optimized since it reduces the merge frequency.
We will discuss the performance of these two merge policies further in \refsection{sec:lsm-cost}.

\subsubsection{Some Well-Known Optimizations}
\label{sec:lsm-optimizations}
There are two well-known optimizations that are used by most LSM-tree implementations today.

\textbf{Bloom Filter.} A Bloom filter~\cite{bloom-filter1970} is a space-efficient probabilistic
data structure designed to aid in answering set membership queries.
It supports two operations, i.e., inserting a key and testing the membership of a given key.
To insert a key, it applies multiple hash functions to map the key into multiple locations in a bit vector
and sets the bits at these locations to 1.
To check the existence of a given key, the key is again hashed to multiple locations.
If all of the bits are 1, then the Bloom filter reports that the key probably exists.
By design, the Bloom filter can report false positives but not false negatives.

Bloom filters can be built on top of disk components to greatly improve point lookup performance.
To search a disk component, a point lookup query can first check its Bloom filter
and then proceed to search its \btree only if its associated Bloom filter reports a positive answer.
Alternatively, a Bloom filter can be built for each leaf page of a disk component.
In this design, a point lookup query can first search the non-leaf pages of a \btree to locate the leaf page, where the non-leaf pages
are assumed to be small enough to be cached, 
and then check the associated Bloom filter before fetching the leaf page to reduce disk I/Os.
Note that the false positives reported by a Bloom filter do not impact the correctness of a query,
but a query may waste some I/O searching for non-existent keys.
The false positive rate of a Bloom filter can be computed as $(1-e^{-kn/m})^k$, where $k$ is the number of hash functions, $n$ is the number of keys, and $m$ is the total number of bits~\cite{bloom-filter1970}.
Furthermore, the optimal number of hash functions that minimizes the false positive rate is $k=\frac{m}{n}ln2$.
In practice, most systems typically use 10 bits/key as a default configuration, which gives a 1\% false positive rate.
Since Bloom filters are very small and can often be cached in memory,
the number of disk I/Os for point lookups is greatly reduced by their use.

\textbf{Partitioning.}
Another commonly adopted optimization is to range-partition the disk components of LSM-trees into multiple
(usually fixed-size) small partitions.
To minimize the potential confusion caused by different terminologies,
we use the term \emph{SSTable} to denote such a partition,
following the terminology from LevelDB~\cite{leveldb}.
This optimization has several advantages.
First, partitioning breaks a large component merge operation into multiple smaller ones,
bounding the processing time of each merge operation as well as the temporary disk space needed to create new components.
Moreover, partitioning can optimize for workloads with sequentially created keys or skewed updates
by only merging components with overlapping key ranges.
For sequentially created keys, essentially no merge is performed since there are no components with overlapping key ranges.
For skewed updates, the merge frequency of the components with cold update ranges can be greatly reduced.
It should be noted that the original LSM-tree~\cite{lsm1996} automatically takes advantage of partitioning because
of its rolling merges.
However, due to the implementation complexity of its rolling merges,
today's LSM-tree implementations typically opt for actual physical partitioning rather than rolling merges.

An early proposal that applied partitioning to LSM-trees is the partitioned exponential file (PE-file)~\cite{pe-file2007}.
A PE-file contains multiple partitions, where each partition can be logically viewed as a separate LSM-tree.
A partition can be further split into two partitions when it becomes too large.
However, this design enforces strict key range boundaries among partitions, which reduces the flexibility of merges.

We now discuss the partitioning optimization used in today's LSM-tree implementations.
It should be noted that partitioning is orthogonal to merge policies;
both leveling and tiering (as well as other emerging merge policies) can be adapted to support partitioning.
To the best of our knowledge, only the partitioned leveling policy has been fully
implemented by industrial LSM-based storage systems,
such as LevelDB~\cite{leveldb} and RocksDB~\cite{rocksdb}.
Some recent papers ~\cite{wb-tree2013,sifrdb2018,pebblesdb2017,lsm-trie2015,lwc-tree2017} have proposed various forms of a partitioned tiering merge policy to achieve better write performance\footnote{RocksDB supports a limited form of a partitioned tiering merge policy
	to bound the maximum size of each SSTable due to its internal restrictions.
	However, the disk space may still be doubled temporarily during large merges.}.

\begin{figure}
	\centering
	\includegraphics[width=0.85\linewidth]{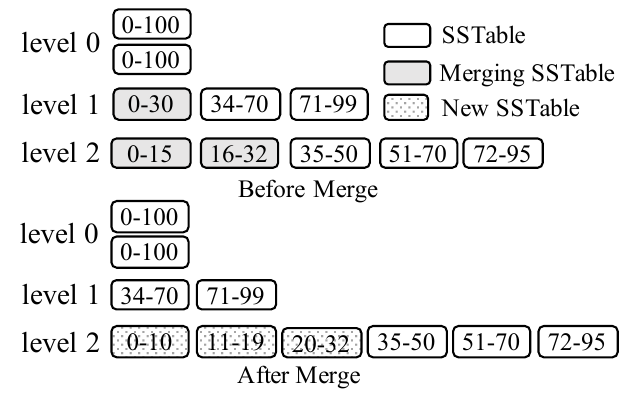}
	\vspace{-0.1in}
	\caption{Partitioned leveling merge policy}
	\label{fig:partitioned-leveling}
\end{figure}

In the partitioned leveling merge policy, pioneered by LevelDB~\cite{leveldb},
the disk component at each level is range-partitioned into multiple fixed-size SSTables, as shown in \reffigure{fig:partitioned-leveling}.
Each SSTable is labeled with its key range in the figure.
Note that the disk components at level 0 are not partitioned since they are directly flushed from memory.
This design can also help the system to absorb write bursts since it can tolerate multiple unpartitioned components at level 0.
To merge an SSTable from level $L$ into level $L+1$, all of its overlapping SSTables at level $L+1$ are selected,
and these SSTables are merged with it to produce new SSTables still at level $L+1$.
For example, in the figure, the SSTable labeled 0-30 at level 1 is merged with the SSTables labeled 0-15 and 16-32 at level 2.
This merge operation produces new SSTables labeled 0-10, 11-19, and 20-32 at level 2,
and the old SSTables will then be garbage-collected.
Different policies can be used to select which SSTable to merge next at each level.
For example, LevelDB uses a round-robin policy (to minimize the total write cost).

\begin{figure}
	\centering
	\includegraphics[width=0.85\linewidth]{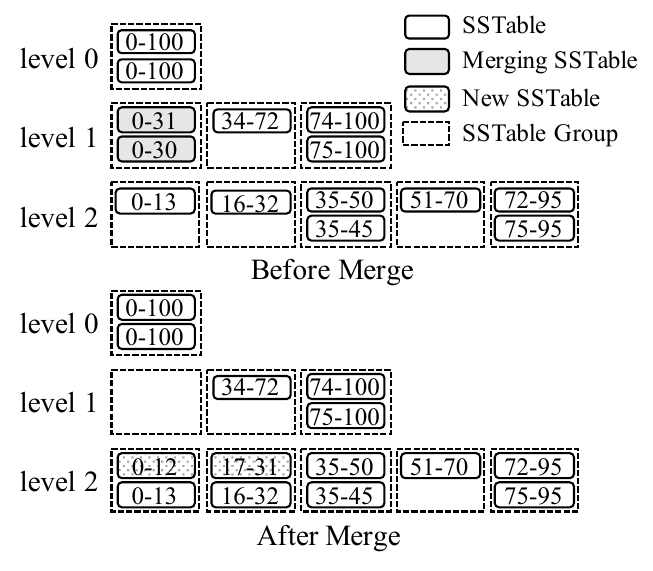}
	\vspace{-0.1in}
	\caption{Partitioned tiering with vertical grouping}
	\label{fig:partitioned-tiering-vertical}
\end{figure}

\begin{figure}
	\includegraphics[width=0.92\linewidth]{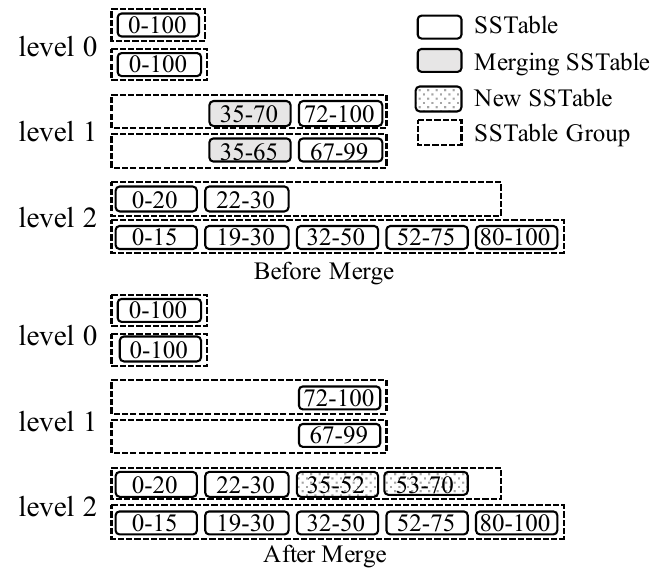}
	\vspace{-0.1in}
	\caption{Partitioned tiering with horizontal grouping}
	\label{fig:partitioned-tiering-horizontal}
\end{figure}

The partitioning optimization can also be applied to the tiering merge policy.
However, one major issue in doing so is that each level can contain multiple SSTables with overlapping key ranges.
These SSTables must be ordered properly based on their recency to ensure correctness.
Two possible schemes can be used to organize the SSTables at each level,
namely vertical grouping and horizontal grouping.
In both schemes, the SSTables at each level are organized into groups.
The vertical grouping scheme groups SSTables with overlapping key ranges together so that the groups have disjoint key ranges.
Thus, it can be viewed as an extension of partitioned leveling to support tiering.
Alternatively, under the horizontal grouping scheme, each logical disk component, which is range-partitioned into a set of SSTables,
serves as a group directly.
This allows a disk component to be formed incrementally based on the unit of SSTables.
We will discuss these two schemes in detail below.

An example of the vertical grouping scheme is shown in \reffigure{fig:partitioned-tiering-vertical}.
In this scheme, SSTables with overlapping key ranges are grouped together so that the groups have disjoint key ranges.
During a merge operation, all of the SSTables in a group are merged together to produce the resulting SSTables based on
the key ranges of the overlapping groups at the next level, which are then added to these overlapping groups.
For example in the figure, the SSTables labeled 0-30 and 0-31 at level 1 are merged together
to produce the SSTables labeled 0-12 and 17-31, which are then added to the overlapping groups at level 2.
Note the difference between the SSTables before and after this merge operation.
Before the merge operation, the SSTables labeled 0-30 and 0-31 have overlapping
key ranges and both must be examined together by a point lookup query.
However, after the merge operation, the SSTables labeled 0-12 and 17-31 have
disjoint key ranges and only one of them needs to be examined by a point lookup query.
It should also be noted that under this scheme
SSTables are no longer fixed-size since they are produced based on
the key ranges of the overlapping groups at the next level.

\reffigure{fig:partitioned-tiering-horizontal} shows an example of the horizontal grouping scheme.
In this scheme, each component, which is range-partitioned into a set of fixed-size SSTables, serves as a logical group directly.
Each level $L$ further maintains an active group,
which is also the first group, to receive new SSTables merged from the previous level.
This active group can be viewed as a partial component being formed by merging
the components at level $L-1$ in the unpartitioned case.
A merge operation selects the SSTables with overlapping key ranges from all of the groups at a level,
and the resulting SSTables are added to the active group at the next level.
For example in the figure, the SSTables labeled 35-70 and 35-65 at level 1 are merged together,
and the resulting SSTables labeled 35-52 and 53-70 are added to the first group at level 2.
However, although SSTables are fixed-size under the horizontal grouping scheme,
it is still possible that one SSTable from a group may overlap a large number of SSTables
in the remaining groups.

\subsubsection{Concurrency Control and Recovery}
We now briefly discuss the concurrency control and recovery techniques used by today's LSM-tree implementations.
For concurrency control, an LSM-tree needs to handle concurrent reads and writes
and to take care of concurrent flush and merge operations.
Ensuring correctness for concurrent reads and writes is a general requirement for access methods in a database system.
Depending on the transactional isolation requirement, today's LSM-tree
implementations either use a locking scheme~\cite{asterixdb2014} or a multi-version scheme~\cite{cassandra,hbase,rocksdb}.
A multi-version scheme works well with an LSM-tree since obsolete versions of a key
can be naturally garbage-collected during merges.
Concurrent flush and merge operations, however, are unique to LSM-trees.
These operations modify the metadata of an LSM-tree, e.g., the list of active components.
Thus, accesses to the component metadata must be properly synchronized.
To prevent a component in use from being deleted, each component can maintain a reference counter.
Before accessing the components of an LSM-tree, a query can first obtain a snapshot of active components
and increment their in-use counters.

Since all writes are first appended into memory, write-ahead logging (WAL) can be performed to ensure their durability.
To simplify the recovery process, existing systems typically employ a no-steal buffer management policy~\cite{recovery1983}.
That is, a memory component can only be flushed when all active write transactions have terminated.
During recovery for an LSM-tree, the transaction log is replayed to redo all successful transactions,
but no undo is needed due to the no-steal policy.
Meanwhile, the list of active disk components must also be recovered in the event of a crash.
For unpartitioned LSM-trees, this can be accomplished by adding a pair of timestamps to each disk component that indicate the range of timestamps of the stored entries.
This timestamp can be simply generated using local wall-clock time or a monotonic sequence number.
To reconstruct the component list, the recovery process can simply find all components with disjoint timestamps.
In the event that multiple components have overlapping timestamps,
the component with the largest timestamp range is chosen and the rest can simply be
deleted since they will have been merged to form the selected component.
For partitioned LSM-trees, this timestamp-based approach does not work anymore since each component is further range-partitioned.
To address this, a typical approach, used in LevelDB~\cite{leveldb} and RocksDB~\cite{rocksdb},
is to maintain a separate metadata log to store all changes to the structural metadata, such as adding or deleting SSTables.
The state of the LSM-tree structure can then be reconstructed by replaying the metadata log during recovery.

\subsection{Cost Analysis}
\label{sec:lsm-cost}
To help understand the performance trade-offs of LSM-trees,
we can turn to the cost analysis of writes, point lookups, range queries, and space amplification
presented in~\cite{dostoevsky2018,monkey2017}.
The cost of writes and queries is measured by counting the number of disk I/Os per operation.
This analysis considers an unpartitioned LSM-tree and represents a worst-case cost.

\begin{table*}
	\caption{Summary of Cost Complexity of LSM-trees}
	\label{table:lsm-cost}
	\centering
	\begin{tabular}{ccP{4cm}cccc}
		\toprule
		Merge Policy & Write & Point Lookup \linebreak (Zero-Result/ Non-Zero-Result)
		& Short Range Query & Long Range Query & Space Amplification \\
		\midrule
		Leveling & $O(T\cdot \frac{L}{B})$ & $O(L\cdot e^{-\frac{M}{N}})$ / $O(1)$ & $O(L)$ & $O(\frac{s}{B})$ & $O(\frac{T+1}{T})$ \\
		Tiering & $O(\frac{L}{B})$ & $O(T\cdot L \cdot e^{-\frac{M}{N}})$ / $O(1)$ & $O(T\cdot L)$ & $O(T\cdot \frac{s}{B})$ & $O(T)$ \\
		\bottomrule
	\end{tabular}
\end{table*}

Let the size ratio of a given LSM-tree be $T$, and suppose the LSM-tree contains $L$ levels.
In practice, for a stable LSM-tree where the volume of inserts equals the volume of deletes, $L$ remains static.
Let $B$ denote the page size, that is, the number of entries that each data page can store,
and let $P$ denote the number of pages of a memory component.
As a result, a memory component will contain at most $B\cdot P$ entries,
and level $i$ ($i \ge 0$ )will contain at most $ T^{i+1} \cdot B\cdot P$ entries.
Given $N$ total entries, the largest level contains approximately $N\cdot \frac{T}{T+1}$
entries since it is $T$ times larger than the previous level.
Thus, the number of levels for $N$ entries
can be approximated as $L = \ceil{\log_T{(\frac{N}{B\cdot P} \cdot \frac{T}{T+1}} )}$.

The write cost, which is also referred to as write amplification in the literature,
measures the amortized I/O cost of inserting an entry into an LSM-tree.
It should be noted that this cost measures the overall I/O cost for this entry to be merged into the largest level
since inserting an entry into memory does not incur any disk I/O.
For leveling, a component at each level will be merged $T-1$ times until it fills up and is pushed to the next level.
For tiering, multiple components at each level are merged only once and are pushed to the next level directly.
Since each disk page contains $B$ entries, the write cost for each entry will be $O(T\cdot \frac{L}{B})$ for leveling
and $O(\frac{L}{B})$ for tiering.

The I/O cost of a query depends on the number of components in an LSM-tree.
Without Bloom filters, the I/O cost of a point lookup will be $O(L)$ for leveling and $O(T\cdot L)$ for tiering.
However, Bloom filters can greatly improve the point lookup cost.
For a zero-result point lookup,
i.e., for a search for a non-existent key, all disk I/Os are caused by Bloom filter false positives.
Suppose all Bloom filters have $M$ bits in total and have the same false positive rate across all levels.
With $N$ total keys, each Bloom filter has a false positive rate of $O(e^{-\frac{M}{N}})$~\cite{bloom-filter1970}.
Thus, the I/O cost of a zero-result point lookup will be $O(L\cdot e^{-\frac{M}{N}})$
for leveling and $O(T\cdot L\cdot e^{-\frac{M}{N}})$ for tiering.
To search for an existing unique key, at least one I/O must be performed to fetch the entry.
Given that in practice the Bloom filter false positive rate is much smaller than 1,
the successful point lookup I/O cost for both leveling and tiering will be $O(1)$.

The I/O cost of a range query depends on the query selectivity.
Let $s$ be the number of unique keys accessed by a range query.
A range query can be considered to be \emph{long} if $\frac{s}{B}>2\cdot L$, otherwise it is \emph{short}~\cite{dostoevsky2018,monkey2017}.
The distinction is that the I/O cost of a long range query will be dominated by the largest level since the largest level contains most of the data.
In contrast, the I/O cost of a short range query will derive (almost)
equally from all levels since the query must issue one I/O to each disk component.
Thus, the I/O cost of a long range query will be $O(\frac{s}{B})$ for leveling and $O(T \cdot \frac{s}{B})$ for tiering.
For a short range query, the I/O cost will be $O(L)$ for leveling and $O(T\cdot L)$ for tiering.

Finally, let us examine the space amplification of an LSM-tree,
which is defined as the overall number of entries divided by the number of unique entries\footnote{The original analysis presented in~\cite{dostoevsky2018,monkey2017} defines the space amplification to be the overall number of obsolete entries divided by
the number of unique entries. We slightly modified the definition to ensure that the space amplification is no less than 1.}.
For leveling, the worst case occurs when all of the data at the first $L-1$ levels,
which contain approximately $\frac{1}{T}$ of the total data, are updates to the entries at the largest level.
Thus, the worst case space amplification for leveling is $O(\frac{T+1}{T})$.
For tiering, the worst case happens when all of the components at the largest level contain exactly the same set of keys.
As a result, the worst case space amplification for tiering will be $O(T)$.
In practice, the space amplification is an important factor to consider when deploying storage systems~\cite{rocksdb-space2017},
as it directly impacts the storage cost for a given workload.

The cost complexity of LSM-trees is summarized in \reftable{table:lsm-cost}.
Note how the size ratio $T$ impacts the performance of leveling and tiering differently.
In general, leveling is optimized for query performance and space utilization by maintaining one component per level.
However, components must be merged more frequently, which will incur a higher write cost by a factor of $T$.
In contrast, tiering is optimized for write performance by maintaining up to $T$ components at each level.
This, however, will decrease query performance and worsen space utilization by a factor of $T$.
As one can see, the LSM-tree is highly tunable.
For example, by changing the merge policy from leveling to tiering, one can greatly improve write performance with only
a small negative impact on point lookup queries due to the Bloom filters.
However, range queries and space utilization will be significantly impacted.
As we proceed to examine the recent literature on improving LSM-trees,
we will see that each makes certain performance trade-offs.
Actually, based on the RUM conjecture~\cite{rum2016}, each access method has to make certain trade-offs
among the read cost (R), update cost (U), and memory or storage cost (M).
It will be important for the reader to keep in mind the cost complexity described here to better
understand the trade-offs made by the proposed improvements.

\section{LSM-tree Improvements}
\label{sec:lsm-improvements}
In this section we present a taxonomy for use in classifying the existing research efforts on improving LSM-trees.
We then provide an in-depth survey of the LSM-tree literature that follows the structure of the proposed taxonomy.

\subsection{A Taxonomy of LSM-tree Improvements}
Despite the popularity of LSM-trees in modern NoSQL systems,
the basic LSM-tree design suffers from various drawbacks and insufficiencies.
We now identify the major issues of the basic LSM-tree design,
and further present a taxonomy of LSM-tree improvements based on these drawbacks.

\textbf{Write Amplification.} 
Even though LSM-trees can provide much better write throughput than in-place update structures such as \btrees
by reducing random I/Os, the leveling merge policy, which has been adopted by modern key-value stores
such as LevelDB~\cite{leveldb} and RocksDB~\cite{rocksdb},
still incurs relatively high write amplification.
High write amplification not only limits the write performance of an LSM-tree
but also reduces the lifespan of SSDs due to frequent disk writes.
A large body of research has been conducted to reduce the write amplification of LSM-trees.

\textbf{Merge Operations.}
Merge operations are critical to the performance of LSM-trees
and must therefore be carefully implemented.
Moreover, merge operations can have negative impacts on the system,
including buffer cache misses after merges and write stalls during large merges.
Several improvements have been proposed to optimize merge operations to address these problems.

\textbf{Hardware.}
In order to maximize performance, LSM-trees must be carefully implemented to fully utilize the underling hardware platforms.
The original LSM-tree has been designed for hard disks, with the goal being reducing random I/Os.
In recent years, new hardware platforms have presented new opportunities for database systems to achieve better performance.
A significant body of recent research has been devoted to improving LSM-trees to fully exploit the underling hardware platforms,
including large memory, multi-core, SSD/NVM, and native storage.

\textbf{Special Workloads.}
In addition to hardware opportunities, certain special workloads
can also be considered to achieve better performance in those use cases.
In this case, the basic LSM-tree implementation must be adapted and customized
to exploit the unique characteristics exhibited by these special workloads.

\begin{figure*}
	\includegraphics[width=\linewidth]{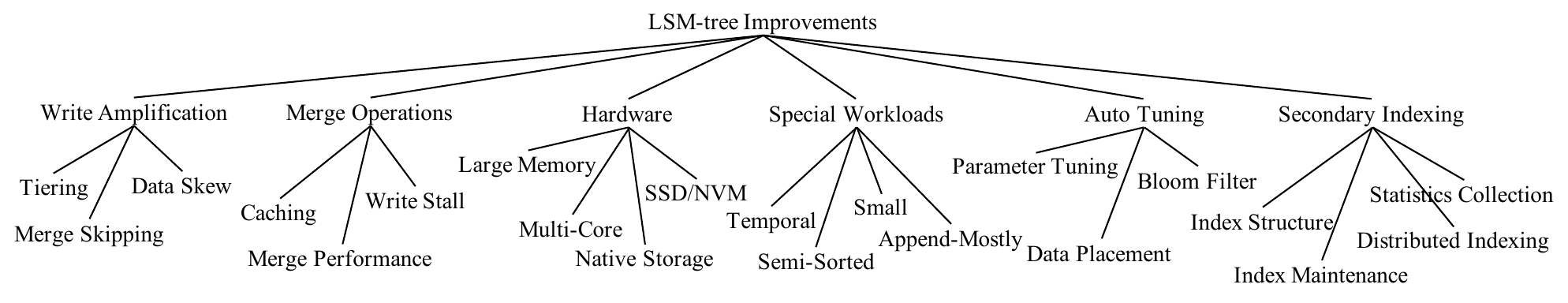}
	\vspace{-0.2in}
	\caption{Taxonomy of LSM-tree improvements}
	\label{fig:taxonomy}
\end{figure*}

\textbf{Auto-Tuning.}
Based on the RUM conjecture~\cite{rum2016}, no access method can be read-optimal, write-optimal, and space-optimal at the same time.
The tunability of LSM-trees is a promising solution to achieve optimal trade-offs for a given workload.
However, LSM-trees can be hard to tune because of too many tuning knobs,
such as memory allocation, merge policy, size ratio, etc.
To address this issue, several auto-tuning techniques have been proposed in the literature.

\textbf{Secondary Indexing.}
A given LSM-tree only provides a simple key-value interface.
To support the efficient processing of queries on non-key attributes, secondary indexes must be maintained.
One issue in this area is how to maintain a set of related secondary indexes efficiently with a small overhead on write performance.
Various LSM-based secondary indexing structures and techniques have been designed and evaluated as well.

Based on these major issues of the basic LSM-tree design,
we present a taxonomy of LSM-tree improvements, shown in \reffigure{fig:taxonomy},
to highlight the specific aspects that the existing research efforts try to optimize.
Given this taxonomy, \reftable{table:category} further classifies the LSM-tree improvements
in terms of each improvement's primary and secondary concerns.
With this taxonomy and classification in hand,
we now proceed to examine each improvement in more depth.

\begin{table*}
	\def\arraystretch{1.05}
	\centering
	\caption{Classification of existing LSM-tree improvements (\primary denotes primary category, \secondary denotes secondary categories)}
	\begin{tabular}{lP{2cm}P{2cm}cP{2cm}P{1.5cm}P{1.5cm}}
		\toprule
		Publication & Write\linebreak Amplification &  Merge\linebreak Operations & Hardware &  Special\linebreak Workloads & Auto\linebreak Tuning & Secondary \linebreak Indexing\\
		\midrule
		WB-tree~\cite{wb-tree2013} & \primary &  & &  & & \\
		LWC-tree~\cite{lwc-tree-tos2017,lwc-tree2017} &\primary  &  &  &  &  & \\
		PebblesDB~\cite{pebblesdb2017} & \primary &   &  &  & & \\
		dCompaction~\cite{dcompaction2017} & \primary &  & &  & \secondary & \\
		Zhang et al.~\cite{groupedLSM2016} & \primary &  & &  & & \\
		SifrDB~\cite{sifrdb2018} & \primary & \secondary & \secondary &  & & \\
		Skip-tree~\cite{skip-tree2017} &\primary &  & &  & & \\
		TRIAD~\cite{triad2017} &\primary &  & &  & & \\
		VT-tree~\cite{vt-tree2013} &  & \primary  & &  & & \\
		Zhang et al.~\cite{pipelined-compaction2014} &  & \primary & &  & & \\
		Ahmad et al.~\cite{compaction2015} &  & \primary & &  & & \\
		LSbM-tree~\cite{lsbm2017,lsbm2018} &  & \primary  & &  & & \\
		bLSM~\cite{blsm2012} &  & \primary  & &  & & \\
		FloDB~\cite{flodb2017} & \secondary  &  & \primary &  & & \\
		Accordion~\cite{accordion2018} & \secondary  &  &\primary &  & & \\
		cLSM~\cite{clsm2015} &  &  &\primary &  & & \\
		FD-tree~\cite{fd-tree2010} &  &  &   \primary &   &  & \\
		FD+tree~\cite{fd+tree2012} &  &\secondary  &   \primary &   & & \\
		MaSM~\cite{masm2011} & \secondary &  &\primary & \secondary & & \\
		WiscKey~\cite{wisckey2017} & \secondary &  &\primary &  & & \\
		HashKV~\cite{hashkv2018} & \secondary &  & \primary &  & & \\
		Kreon~\cite{kreon2018} &\secondary  &  &\primary &  & & \\
		NoveLSM~\cite{novelsm2018} &  &  &\primary &  & & \\
		LDS~\cite{lds2017} &  &  &\primary &  & & \\
		LOCS~\cite{open-channel-ssd} &  &  &  \primary &  & & \\
		NoFTL-KV~\cite{noftl-kv2018} &  &  &\primary &  & & \\
		LHAM~\cite{lham2000} &  &  & & \primary & & \\
		LSM-trie~\cite{lsm-trie2015} &\secondary  &  & &\primary  & & \\
		SlimDB~\cite{slimdb2017} & \secondary &  & &  \primary& & \\
		Mathieu et al.~\cite{stack-merge2014} & \secondary &  & &  \primary& & \\
		Lim et al.~\cite{lsm-model2016} &  &  & & & \primary & \\
		Monkey~\cite{monkey2017,monkey-tods} &  &  & &  &\primary & \\
		Dostoevsky~\cite{dostoevsky2018} & \secondary  &  & & & \primary & \\
		Thonangi and Yang~\cite{partial-merge2017} &\secondary  &  & &  & \primary & \\
		ElasticBF~\cite{elastic-bf2018} &  &  & &  &\primary & \\
		Mutant~\cite{mutant2018} &  &  & &  &\primary & \\
		LSII~\cite{lsii2013} &  &  & &  & &\primary \\
		Kim et al.~\cite{lsm-spatial2017} &  &  & &  & &\primary \\
		Filter~\cite{asterixdb-filter2015} &  &  & &  & &\primary \\
		Qader et al.~\cite{secondary2018} &  &  & &  & &\primary \\
		Diff-Index~\cite{diff-index2014} &  &  & &  & &\primary \\
		DELI~\cite{deli2015}, &  &  & &  & &\primary \\
		Luo and Carey~\cite{extended} &  &  & &  & &\primary \\
		Ildar et al.~\cite{lsm-cardinality2018} &  &  & &  & &\primary \\
		Joseph et al.~\cite{local-global2017} &  &  & &  & &\primary \\
		Zhu et al.~\cite{bulkload2017} &  &  & &  & &\primary \\
		Duan et al.~\cite{view2018}&  &  & &  & &\primary \\
		\bottomrule
	\end{tabular}
	\label{table:category}
\end{table*}

\subsection{Reducing Write Amplification}
In this section, we review the improvements in the literature that aim to reduce the write amplification of LSM-trees.
Most of these improvements are based on tiering since it has much better write performance than leveling.
Other proposed improvements have developed new techniques to perform merge skipping or to exploit data skews.

\subsubsection{Tiering}
One way to optimize write amplification is to apply tiering
since it has much lower write amplification than leveling.
However, recall from \refsection{sec:lsm-cost} that this will lead to worse query performance and space utilization.
The improvements in this category can all be viewed as some variants of the partitioned tiering design
with vertical or horizontal grouping discussed in \refsection{sec:lsm-optimizations}.
Here we will mainly discuss the modifications made by these improvements.

The WriteBuffer (WB) Tree~\cite{wb-tree2013} can be viewed as a variant of the partitioned tiering design with vertical grouping.
It has made the following modifications.
First, it relies on hash-partitioning to achieve workload balance so that each SSTable group roughly stores the same amount of data.
Furthermore, it organizes SSTable groups into a \btree-like structure to enable self-balancing to minimize the total number of levels.
Specifically, each SSTable group is treated like a node in a \btree.
When a non-leaf node becomes full with $T$ SSTables,
these $T$ SSTables are merged together to form new SSTables that are added into its child nodes.
When a leaf node becomes full with $T$ SSTables,
it is split into two leaf nodes by merging all of its SSTables into two leaf nodes with smaller key ranges
so that each new node receives about $T/2$ SSTables.

The light-weight compaction tree (LWC-tree)~\cite{lwc-tree-tos2017,lwc-tree2017}
adopts a similar partitioned tiering design with vertical grouping.
It further presents a method to achieve workload balancing of SSTable groups.
Recall that under the vertical grouping scheme, SSTables are no longer strictly fixed-size since they are produced based on
the key ranges of the overlapping groups at the next level instead of based on their sizes.
In the LWC-tree, if a group contains too many entries, it will shrink the key range of this group after the group has been merged (now temporarily empty) and will widen the key ranges of its sibling groups accordingly.

PebblesDB~\cite{pebblesdb2017} also adopts a partitioned tiering design with vertical grouping.
The major difference is that it determines the key ranges of SSTable groups using the idea of
guards as inspired by the skip-list~\cite{skip-list1990}.
Guards, which are the key ranges of SSTable groups, are selected probabilistically based on inserted keys to achieve workload balance.
Once a guard is selected, it is applied lazily during the next merge.
PebblesDB further performs parallel seeks of SSTables to improve range query performance.

dCompaction~\cite{dcompaction2017} introduces the concept of virtual SSTables and virtual merges to reduce the merge frequency.
A virtual merge operation produces a virtual SSTable that simply points to the input SSTables without performing actual merges.
However, since a virtual SSTable points to multiple SSTables with overlapping ranges, query performance will degrade.
To address this, dCompaction introduces a threshold based on the number of real SSTables to trigger actual merges.
It also lets queries trigger actual merges if a virtual SSTable pointing to too many SSTables is encountered during query processing.
In general, dCompaction delays a merge operation until multiple SSTables can be merged together, and thus it can also be viewed as a variant of the tiering merge policy.

As one can see, the four structures described above all share a similar high-level design based on partitioned tiering with vertical grouping.
They mainly differ in how workload balancing of SSTable groups is performed.
For example, the WB-tree~\cite{wb-tree2013} relies on hashing, but doing so gives up the ability of supporting range queries.
The LWC-tree~\cite{lwc-tree-tos2017,lwc-tree2017} dynamically shrinks the key ranges of dense SSTable groups,
while PebblesDB~\cite{pebblesdb2017} relies on probabilistically selected guards.
In contrast, dCompaction~\cite{dcompaction2017} offers no built-in support for workload balancing.
It is not clear how skewed SSTable groups would impact the performance of these structures,
and future research is needed to understand this problem and evaluate
these workload balancing strategies.

The partitioned tiering design with horizontal grouping
has been adopted by Zhang et al.~\cite{groupedLSM2016} and SifrDB~\cite{sifrdb2018}.
SifrDB also proposes an early-cleaning technique to reduce disk space utilization during merges.
During a merge operation, SifrDB incrementally activates newly produced SSTables and deactivates the old SSTables.
SifrDB further exploits I/O parallelism to speedup query performance by examining multiple SSTables in parallel.

\subsubsection{Merge Skipping}
\begin{figure}
	\centering
	\includegraphics[width=0.8\linewidth]{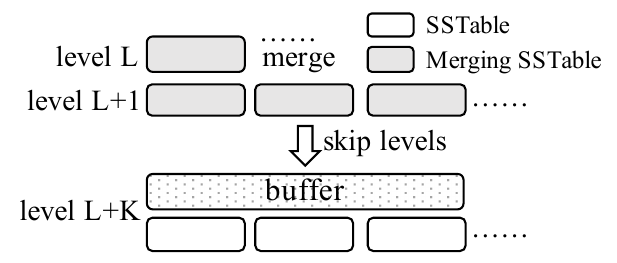}
	\caption{Merge in skip-tree: entries from a lower level can be directly pushed to the mutable buffer of a higher level}
	\label{fig:skip-tree}
\end{figure}

The skip-tree~\cite{skip-tree2017} proposes a merge skipping idea to improve write performance.
The observation is that each entry must be merged from level $0$ down to the largest level.
If some entries can be directly pushed to a higher level by skipping some level-by-level merges,
then the total write cost will be reduced.
As shown in \reffigure{fig:skip-tree}, during a merge at level $L$,
the skip-tree directly pushes some keys to a mutable buffer at level $L+K$
so that some level-by-level merges can be skipped.
Meanwhile, the skipped entries in the mutable buffer will be merged with the SSTables at level $L+K$ during subsequent merges.
To ensure correctness, a key from level $L$ can be pushed to level $L+K$
only if this key does not appear in any of the intermediate levels $L+1,\cdots, L+K-1$.
This condition can be tested efficiently by checking the Bloom filters of the intermediate levels.
The skip-tree further performs write-ahead logging to ensure durability of the entries stored in the mutable buffer.
To reduce the logging overhead,
the skip-tree only logs the key plus the ID of the original SSTable and prevents an SSTable from being deleted
if it is referenced by any key in the buffer.
Although merge skipping is an interesting idea to reduce write amplification,
it introduces non-trivial implementation complexity to manage the mutable buffers.
Moreover, since merge skipping essentially reduces some merges at the intermediate levels,
it is not clear how the skip-tree would compare against a well-tuned LSM-tree by reducing the size ratio.

\subsubsection{Exploiting Data Skew}
TRIAD~\cite{triad2017} reduces write amplification for skewed update workloads where some hot keys are updated frequently.
The basic idea is to separate hot keys from cold keys in the memory component so that only cold keys are flushed to disk.
As a result, when hot keys are updated, old versions can be discarded directly without writing them to disk.
Even though hot keys are not flushed to disk, they are periodically copied to a new transaction log
so that the old transaction log can be reclaimed.
TRIAD also reduces write amplification by delaying merges at level 0 until level 0 contains multiple SSTables.
Finally, it presents an optimization that avoids creating new disk components after flushes.
Instead, the transaction log itself is used as a disk component
and an index structure is built on top of it to improve lookup performance.
However, range query performance will still be negatively impacted since entries are not sorted in the log.

\subsubsection{Summary}
Tiering has been widely used to improve the write performance of LSM-trees,
but this will decrease query performance and space utilization, as discussed in \refsection{sec:lsm-cost}.
The existing tiering-based improvements mainly differ in how SSTables are managed,
either by vertical grouping~\cite{wb-tree2013,dcompaction2017,pebblesdb2017,lwc-tree-tos2017,lwc-tree2017}
or horizontal grouping~\cite{sifrdb2018,groupedLSM2016}.
It is not clear how these different grouping schemes impact system performance
and it would be useful as future work to study and evaluate their impact.
The skip-tree~\cite{skip-tree2017} and TRIAD~\cite{triad2017} propose several
new ideas to improve write performance, ideas that are orthogonal to tiering.
However, these optimizations bring non-trivial implementation complexity to real systems,
such as the mutable buffers introduced by the skip-tree and the use of transaction logs as flushed components by TRIAD.

All of the improvements in this category, as well as some improvements in the later sections,
have claimed that they can greatly improve the write performance of LSM-trees,
but their performance evaluations have often failed to consider the tunability of LSM-trees.
That is, these improvements have mainly been evaluated against a default (untuned) configuration of LevelDB or RocksDB, 
which use the leveling merge policy with a size ratio of 10.
It is not clear how these improvements would compare against well-tuned LSM-trees.
To address this, one possible solution would be to tune RocksDB to achieve a similar write throughput
to the proposed improvements by changing the size ratio or by adopting the tiering merge policy
and then evaluating how these improvements can improve query performance and space amplification.
Moreover, these improvements have primarily focused on query performance;
space amplification has often been neglected.
It would be a useful experimental study to fully evaluate these improvements against
well-tuned baseline LSM-trees to evaluate their actual usefulness.
We also hope that this situation can be avoided in future research by considering the tunability of LSM-trees
when evaluating the proposed improvements.

\subsection{Optimizing Merge Operations}
Next we review some existing work
that improves the implementation of merge operations, including improving merge performance,
minimizing buffer cache misses, and eliminating write stalls.

\subsubsection{Improving Merge Performance}
The VT-tree~\cite{vt-tree2013} presents a stitching operation to improve merge performance.
The basic idea is that when merging multiple SSTables,
if the key range of a page from an input SSTable does not overlap the key ranges of any pages from other SSTables,
then this page can be simply pointed to by the resulting SSTable without reading and copying it again.
Even though stitching improves merge performance for certain workloads, it has a number of drawbacks.
First, it can cause fragmentation since pages are no longer continuously stored on disk.
To alleviate this problem, the VT-tree introduces a stitching threshold $K$ so that a stitching operation is triggered only
when there are at least $K$ continuous pages from an input SSTable.
Moreover, since the keys in stitched pages are not scanned during a merge operation, 
a Bloom filter cannot be produced.
To address this issue, the VT-tree uses quotient filters~\cite{quotient-filter2012} since multiple quotient filters can be combined directly without accessing the original keys.

\begin{figure}[b]
	\centering
	\includegraphics[width=\linewidth]{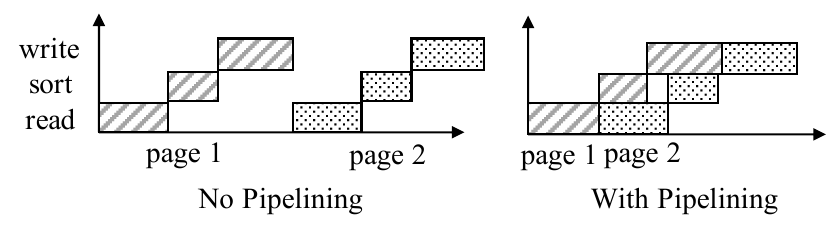}
	\vspace{-0.2in}
	\caption{Pipelined merge example: multiple input pages can be processed in a pipelined fashion}
	\label{fig:pipeline}
\end{figure}

Zhang et al.~\cite{pipelined-compaction2014} proposed a pipelined merge implementation to better utilize
CPU and I/O parallelism to improve merge performance.
The key observation is that a merge operation contains multiple phases, including the read phase, merge-sort phase, and write phase.
The read phase reads pages from input SSTables,
which will then be merge-sorted to produce new pages during the merge-sort phase.
Finally, the new pages will be written to disk during the write phase.
Thus, the read phase and write phase are I/O heavy while the merge-sort phase is CPU heavy.
To better utilize CPU and I/O parallelism, the proposed approach pipelines
the execution of these three phases, as illustrated by \reffigure{fig:pipeline}.
In this example, after the first input page has been read, this approach continues reading the second input page (using disk)
and the first page can be merge-sorted (using CPU).

\subsubsection{Reducing Buffer Cache Misses}
Merge operations can interfere with the caching behavior of a system.
After a new component is enabled,
queries may experience a large number of buffer cache misses since the new component has not been cached yet.
A simple write-through cache maintenance policy cannot solve this problem.
If all of the pages of the new component were cached during a merge operation, a lot of other working pages would be evicted,
which will again cause buffer cache misses.

Ahmad et al.~\cite{compaction2015} conducted an experimental study of the impact of merge operations on system performance.
They found that merge operations consume a large number of CPU and I/O resources and impose a high overhead on query response times.
To address this, this work proposed to offload large merges to remote servers to minimize their impact.
After a merge operation is completed, a smart cache warmup algorithm is used to
fetch the new component incrementally to minimize buffer cache misses.
The idea is to switch to the new component incrementally, chunk by chunk, to smoothly redirect incoming queries
from the old components to the new component.
As a result, the burst of buffer cache misses is decomposed into a large number of smaller ones, minimizing the negative impact of component switching on query performance.

\begin{figure}
	\centering
	\includegraphics[width=0.9\linewidth]{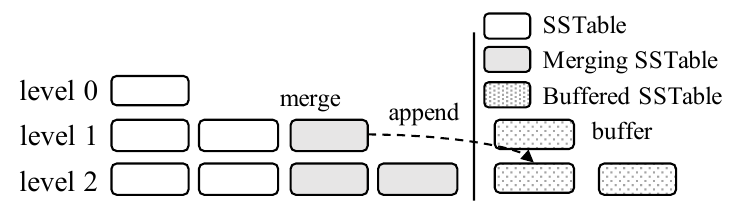}
	\caption{LSbM-tree: the merged component is added to a buffer of the next level instead of being deleted immediately}
	\label{fig:lsbm-tree}
\end{figure}

One limitation of the approach proposed by Ahmad et al.~\cite{compaction2015}
is that merge operations must be offloaded to separate servers.
The incremental warmup algorithm alone was subsequently found to be insufficient
due to contention between the newly produced pages and the existing hot pages~\cite{lsbm2017,lsbm2018}.
To address this limitation, the Log-Structured buffered Merge tree (LSbM-tree) \cite{lsbm2017,lsbm2018}
proposes an alternative approach.
As illustrated by \reffigure{fig:lsbm-tree}, after an SSTable at level $L$ is merged into level $L+1$,
the old SSTables at level $L$ is appended to a buffer associated with level $L+1$ instead of being deleted immediately.
Note that there is no need to add old SSTables at level $L+1$ into the buffer, as the SSTables at $L+1$ all come from level $L$ and the entries of these old SSTables will have already been added to the buffer before.
The buffered SSTables are searched by queries as well to minimize buffer cache misses,
and they are deleted gradually based on their access frequency.
This approach does not incur any extra disk I/O during a merge operation since it only delays the deletion of the old SSTables.
However, this approach is mainly effective for skewed workloads where only a small range of keys are frequently accessed.
It can introduce extra overhead for queries accessing cold data that are not cached,
especially for range queries since they cannot benefit from Bloom filters.

\subsubsection{Minimizing Write Stalls}
Although the LSM-tree offers a much higher write throughput compared to traditional \btrees,
it often exhibits write stalls and unpredictable write latencies
since heavy operations such as flushes and merges run in the background.
bLSM~\cite{blsm2012} proposes a spring-and-gear merge scheduler to minimize write stalls for
the unpartitioned leveling merge policy.
Its basic idea is to tolerate an extra component at each level so that merges at different levels can proceed in parallel.
Furthermore, the merge scheduler controls the progress of merge operations
to ensure that level $L$ produces a new component at level $L+1$ only after the previous merge operation at level $L+1$ has completed.
This eventually cascades to limit the maximum write speed at the memory component and eliminates large write stalls.
However, bLSM itself has several limitations.
bLSM was only designed for the unpartitioned leveling merge policy.
Moreover, it only bounds the maximum latency of writing to memory components
while the queuing latency, which is often a major source of performance variability,
is ignored.

\subsubsection{Summary}
The improvements in this category optimize the implementation of merge operations
in terms of performance, buffer cache misses, and write stalls.
To speedup merge operations, the VT-tree~\cite{vt-tree2013} introduces the stitching operation that avoids copying input pages if applicable.
However, this may cause fragmentation, which is undesirable for hard disks.
Moreover, this optimization is incompatible with Bloom filters, which are widely used in modern LSM-tree implementations.
The pipelined merge implementation~\cite{pipelined-compaction2014} improves merge performance by exploiting CPU and I/O parallelism.
It should be noted that many LSM-based storage systems have already implemented some form of pipelining by exploiting disk read-ahead and write-behind.

Ahmed et al.~\cite{compaction2015} and the LSbM-tree~\cite{lsbm2017,lsbm2018}
present two alternative methods to alleviating buffer cache misses caused by merges.
However, both approaches appear to have certain limitations.
The approach proposed by Ahmed et al.~\cite{compaction2015} requires dedicated servers to perform merges,
while the LSbM-tree~\cite{lsbm2017,lsbm2018} delays the deletion of old components
that could negatively impact queries accessing cold data.
Write stalls are a unique problem of LSM-trees due to its out-of-place update nature.
bLSM~\cite{blsm2012} is the only effort that attempts to address this problem.
However, bLSM~\cite{blsm2012} only bounds the maximum latency of writing to memory components.
The end-to-end write latency can still exhibit large variances due to queuing.
More efforts need to be done to improve the performance stability of LSM-trees.

\subsection{Hardware Opportunities}
We now review the LSM-tree improvements proposed for different hardware platforms,
including large memory, multi-core, SSD/NVM, and native storage.
A general paradigm of these improvements is to modify the basic design of LSM-trees to fully exploit the unique features provided
by the target hardware platform to achieve better performance.

\subsubsection{Large Memory}
It is beneficial for LSM-trees to have large memory components to reduce the total number of levels,
as this will improve both write performance and query performance.
However, managing large memory components brings several new challenges.
If a memory component is implemented directly using on-heap data structures,
large memory can result in a large number of small objects that lead to significant GC overheads.
In contrast, if a memory component is implemented using off-heap structures such as a concurrent \btree,
large memory can still cause a higher search cost (due to tree height)
and cause more CPU cache misses for writes,
as a write must first search for its position in the structure.

FloDB~\cite{flodb2017} presents a two-layer design to manage large memory components.
The top level is a small concurrent hash table to support fast writes,
and the bottom level is a large skip-list to support range queries efficiently.
When the hash table is full, its entries are efficiently migrated into the skip-list using a batched algorithm.
By limiting random writes to a small memory area, this design significantly improves the in-memory write throughput.
To support range queries, FloDB requires that a range query must wait for the hash table to be drained
so that the skip-list alone can be searched to answer the query.
However, FloDB suffers from two major problems.
First, it is not efficient for workloads containing both writes and range queries due to their contention.
Second, the skip-list may have a large memory footprint and lead to lower memory utilization.

\begin{figure}
	\centering
	\includegraphics[width=0.65\linewidth]{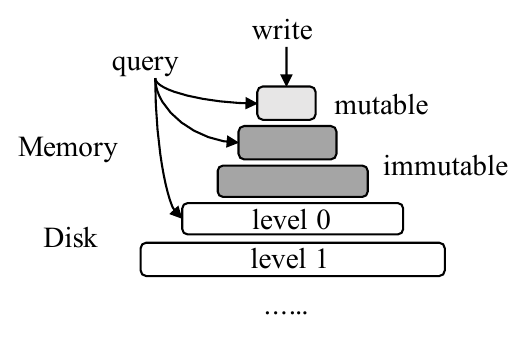}
	\vspace{-0.1in}
	\caption{Accordion's multi-layer structure}
	\label{fig:accordion}
\end{figure}

To address the drawbacks of FloDB, Accordion~\cite{accordion2018}
uses a multi-layer approach to manage its large memory components.
In this design (\reffigure{fig:accordion}), there is a small mutable memory component
in the top level to process writes.
When the mutable memory component is full, instead of being flushed to disk,
it is simply flushed into a (more compact) immutable memory component via an in-memory flush operation.
Similarly, such immutable memory components can be merged via in-memory merge operations to
improve query performance and reclaim space occupied by obsolete entries.
Note that in-memory flush and merge operations do not involve any disk I/O,
which reduces the overall disk I/O cost by leveraging large memory.

\subsubsection{Multi-Core}
cLSM~\cite{clsm2015} optimizes for multi-core machines and presents new concurrency control algorithms for various LSM-tree operations.
It organizes LSM components into a concurrent linked list to minimize blocking caused by synchronization.
Flush and merge operations are carefully designed so that they only result in atomic modifications to the linked list that will never block queries.
When a memory component becomes full, a new memory component is allocated while the old one will be flushed.
To avoid writers inserting into the old memory component, 
a writer acquires a shared lock before modifications and the flush thread acquires an exclusive lock before flushes.
cLSM also supports snapshot scans via multi-versioning and atomic read-modify-write operations
using an optimistic concurrency control approach that exploits the fact that all writes, and thus all conflicts, involve the memory component.

\subsubsection{SSD/NVM}
Different from traditional hard disks, which only support efficient sequential I/Os,
new storage devices such as solid-state drives (SSDs) and non-volatile memories (NVMs) support efficient random I/Os as well.
NVMs further provide efficient byte-addressable random accesses with persistence guarantees.

\begin{figure}
	\centering
	\includegraphics[width=\linewidth]{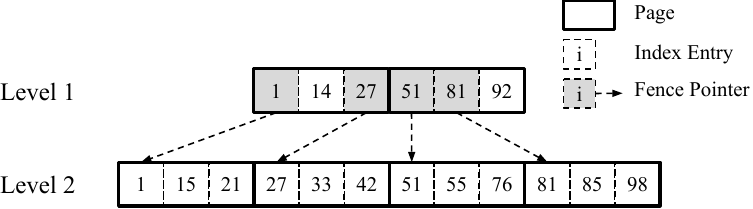}
	\caption{Example FD-tree structure}
	\label{fig:fd-tree}
\end{figure}

The FD-tree~\cite{fd-tree2010} uses a similar design to LSM-trees to reduce random writes on SSDs.
One major difference is that the FD-tree exploits fractional cascading~\cite{fractional-cascading1986}
to improve query performance instead of Bloom filters.
For the component at each level, the FD-tree additionally stores fence pointers
that point to each page at the next level.
For example in \reffigure{fig:fd-tree}, the pages at level $2$ are pointed at by fence pointers with keys 1, 27, 51, 81 at level 1.
After performing a binary search at level 0, a query can follow these fence pointers to traverse all of the levels.
However, this design introduces additional complexity to merges.
When the component at level $L$
is merged into level $L+1$, all of the previous levels $0$ to $L-1$ must be merged as well to rebuild the fence pointers.
Moreover, a point lookup still needs to perform disk I/Os when searching for non-existent keys, which can be mostly avoided by using Bloom filters.
For these reasons, modern LSM-tree implementations prefer Bloom filters rather than fractional cascading\footnote{RocksDB~\cite{rocksdb}
supports a limited form of fractional cascading by maintaining the set of overlapping SSTables at the adjacent next level for each SSTable.
These pointers are used to narrow down the search range when locating specific SSTables during point lookups.}. 

The FD+tree~\cite{fd+tree2012} improves the merge process of the FD-tree~\cite{fd-tree2010}.
In the FD-tree, when a merge happens from level $0$ to level $L$, new components at levels $0$ to $L$ must be created,
which will temporarily double the disk space.
To address this, during a merge operation, the FD+tree incrementally activates the new components
and reclaims pages from the old components that are not used by any active queries.

MaSM (materialized sort-merge)~\cite{masm2011} is designed for supporting efficient updates
for data warehousing workloads by exploiting SSDs.
MaSM first buffers all updates into an SSD.
It uses the tiering merge policy to merge intermediate components with low write amplification.
The updates are then merged back to the base data, which resides in the hard disk.
MaSM can be viewed as a simplified form of the lazy leveling merge policy proposed by
Dostoevsky~\cite{dostoevsky2018}, as we will see later in this survey.
Moreover, since MaSM mainly targets long range queries to support data warehousing workloads,
the overhead introduced by intermediate components stored in SSDs is negligible compared to the cost of accessing the base data.
This enables MaSM to only incur a small overhead on queries with concurrent updates.

\begin{figure}[b]
	\centering
	\includegraphics[width=\linewidth]{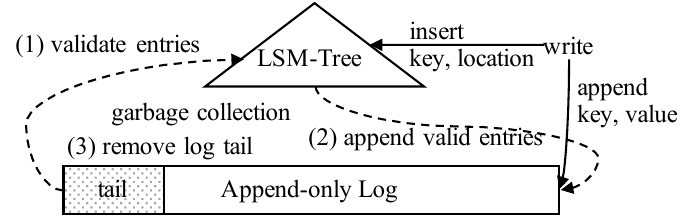}
	\caption{WiscKey stores values into an append-only log to reduce the write amplification of the LSM-tree}
	\label{fig:wisckey}
\end{figure}

Since SSDs support efficient random reads, separating values from keys
becomes a viable solution to improve the write performance of LSM-trees.
This approach was first implemented by WiscKey~\cite{wisckey2017}
and subsequently adopted by HashKV~\cite{hashkv2018} and SifrDB~\cite{sifrdb2018}.
As shown in \reffigure{fig:wisckey}, WiscKey~\cite{wisckey2017} stores key-value pairs into an append-only log
and the LSM-tree simply serves as a primary index that maps each key to its location in the log.
While this can greatly reduce the write cost by only merging keys,
range query performance will be significantly impacted because values are not sorted anymore.
Moreover, the value log must be garbage-collected efficiently to reclaim the storage space.
In WiscKey, garbage-collection is performed in three steps.
First, WiscKey scans the log tail and validates each entry by performing point lookups against the LSM-tree to find out whether the location of each key has changed or not.
Second, valid entries, whose locations have not changed, are then appended to the log and their locations are updated in the LSM-tree as well.
Finally, the log tail is truncated to reclaim the storage space.
However, this garbage-collection process has been shown to be a new performance bottleneck~\cite{hashkv2018}
due to its expensive random point lookups.

HashKV~\cite{hashkv2018} introduces a more efficient approach to garbage-collect obsolete values.
The basic idea is to hash-partition the value log into multiple partitions based on keys
and to garbage-collect each partition independently.
In order to garbage-collect a partition, HashKV performs a group-by operation on the keys to find the latest value for each key.
Valid key-value pairs are added to a new log and their locations are then updated in the LSM-tree.
HashKV further stores cold entries separately so that they can be garbage-collected less frequently.

Kreon~\cite{kreon2018} exploits memory-mapped I/O to reduce CPU overhead by avoiding unnecessary data copying.
It implements a customized memory-mapped I/O manager in the Linux kernel
to control cache replacement and to enable blind writes.
To improve range query performance, Kreon reorganizes data during query processing
by storing the accessed key-value pairs together in a new place.

NoveLSM~\cite{novelsm2018} is an implementation of LSM-trees on NVMs.
NoveLSM adds an NVM-based memory component to serve writes when the DRAM memory component
is full so that writes can still proceed without being stalled.
It further optimizes the write performance of the NVM memory component by skipping logging
since NVM itself provides persistence.
Finally, it exploits I/O parallelism to search multiple levels concurrently to reduce lookup latency.

\subsubsection{Native Storage}
Finally, the last line of work in this category attempts to perform native management of storage devices, such as HDDs and SSDs,
to optimize the performance of LSM-tree implementations.

The LSM-tree-based Direct Storage system (LDS)~\cite{lds2017} bypasses the file system to
better exploit the sequential
and aggregated I/O patterns exhibited by LSM-trees.
The on-disk layout of LDS contains three parts: chunks, a version log, and a backup log.
Chunks store the disk components of the LSM-tree.
The version log stores the metadata changes of the LSM-tree after each flush and merge.
For example, a version log record can record the obsolete chunks and the new chunks resulting from a merge.
The version log is regularly checkpointed to aggregate all changes so that the log can be truncated.
Finally, the backup log provides durability for in-memory writes by write-ahead logging.

LOCS~\cite{open-channel-ssd} is an implementation of the LSM-tree on open-channel SSDs.
Open-channel SSDs expose internal I/O parallelism via an interface called channels,
where each channel functions independently as a logical disk device.
This allows applications to flexibly schedule disk writes to leverage the available I/O parallelism,
but disk reads must be served by the same channel where the data is stored.
To exploit this feature, LOCS dispatches disk writes due to flushes and merges
to all channels using a least-weighted-queue-length policy
to balance the total amount of work allocated to each channel.
To further improve the I/O parallelism for partitioned LSM-trees,
LOCS places SSTables from different levels with similar key ranges into different channels
so that these SSTables can be read in parallel.

NoFTL-KV~\cite{noftl-kv2018} proposes to extract the flash translation layer (FTL) from the storage device into the key-value store to gain direct control over the storage device.
Traditionally, the FTL translates the logical block address to the physical block address to implement wear leveling, which improves the lifespan of SSDs by distributing writes evenly to all blocks.
NoFTL-KV argues for a number of advantages of extracting FTL,
such as pushing tasks down to the storage device, performing more efficient data placement to exploit I/O parallelism,
and integrating the garbage-collection process of the storage device with the merge process of LSM-trees
to reduce write amplification.

\subsubsection{Summary}
In this subsection, we have reviewed the LSM-tree improvements exploiting hardware platforms,
including large memory~\cite{flodb2017,accordion2018},
multi-core~\cite{clsm2015},
SSD/NVM~\cite{masm2011,hashkv2018,novelsm2018,fd-tree2010,wisckey2017,kreon2018,fd+tree2012},
and native storage~\cite{lds2017,open-channel-ssd,noftl-kv2018}.
To manage large memory components, both FloDB~\cite{flodb2017} and Accordion~\cite{accordion2018}
take a multi-layer approach to limit random writes to a small memory area.
The difference is that FloDB~\cite{flodb2017} only uses two layers, while Accordion~\cite{accordion2018}
uses multiple layers to provide better concurrency and memory utilization.
For multi-core machines, cLSM~\cite{clsm2015} presents a set of new concurrency control algorithms to improve concurrency.

A general theme of the improvements for SSD/NVM is to exploit the high random read throughput
while reducing the write amplification of LSM-trees to improve the lifespan of these storage devices.
The FD-tree~\cite{fd-tree2010} and its successor FD+tree~\cite{fd+tree2012}
propose to use fractional cascading~\cite{fractional-cascading1986} to improve point lookup performance
so that only one random I/O is needed for searching each component.
However, today's implementations generally prefer Bloom filters since unnecessary I/Os can be mostly avoided by point lookups.
Separating keys from values~\cite{hashkv2018,wisckey2017,kreon2018} can significantly improve
the write performance of LSM-trees since only keys are merged.
However, this leads to lower query performance and space utilization.
Meanwhile, values must be garbage-collected separately to reclaim disk space,
which is similar to the traditional log-structured file system design~\cite{lfs1992}.
Finally, some recent work has proposed to perform native management of storage devices, including HDDs~\cite{lds2017}
and SSDs~\cite{noftl-kv2018,open-channel-ssd},
which can often bring large performance gains by exploiting the sequential and non-overwriting
I/O patterns exhibited by LSM-trees.

\subsection{Handling Special Workloads}
We now review some existing LSM-tree improvements that target special workloads to achieve better performance.
The considered special workloads include temporal data, small data, semi-sorted data, and append-mostly data.

The log-structured history access method (LHAM) \cite{lham2000}
improves the original LSM-tree to more efficiently support temporal workloads.
The key improvement made by LHAM is to attach a range of timestamps
to each component to facilitate the processing of temporal queries by pruning irrelevant components.
It further guarantees that the timestamp ranges of components are disjoint from one another.
This is accomplished by modifying the rolling merge process
to always merge the records with the oldest timestamps from a component $C_{i}$ into $C_{i+1}$.

\begin{figure}
	\centering
	\includegraphics[width=\linewidth]{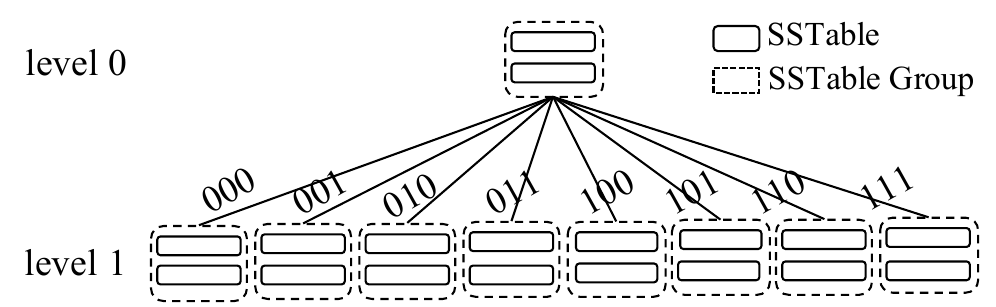}
	\caption{The LSM-trie uses the prefix of hash values to manage SSTables. In this example, each level uses three bits to perform partitioning.}
	\label{fig:lsm-trie}
\end{figure}

The LSM-trie~\cite{lsm-trie2015} is an LSM-based hash index for managing a large number of key-value pairs
where each key-value pair is small.
It proposes a number of optimizations to reduce the metadata overhead.
The LSM-trie adopts a partitioned tiering design to reduce write amplification.
Instead of storing the key ranges of each SSTable directly, the LSM-trie organizes its SSTables using the prefix of their
hash values to reduce the metadata overhead, as shown in \reffigure{fig:lsm-trie}.
The LSM-trie further eliminates the index page,
instead assigning key-value pairs into fixed-size buckets based on their hash values.
Overflow key-value pairs are assigned to underflow buckets and this information is recorded in a migration metadata table.
The LSM-trie also builds a Bloom filter for each bucket.
Since there are multiple SSTables in each group at a level, the LSM-trie clusters all Bloom filters
of the same logical bucket of these SSTables together so that they can be fetched using a single I/O by a point lookup query.
In general, the LSM-trie is mainly effective when the number of key-value pairs is so large that
even the metadata, e.g., index pages and Bloom filters, cannot be totally cached.
However, the LSM-trie only supports point lookups since its optimizations heavily depend on hashing.

SlimDB~\cite{slimdb2017} targets semi-sorted data in which each key contains a prefix x and a suffix y.
It supports normal point lookups, given both the prefix and the suffix,
as well as retrieving all the key-values pairs sharing the same prefix key x.
To reduce write amplification, SlimDB adopts a hybrid structure with tiering on the lower levels and leveling on the higher levels.
SlimDB further uses multi-level cuckoo filters~\cite{cuckoo-filter2014}
to improve point lookup performance for levels that use the tiering merge policy.
At each level, a multi-level cuckoo filter maps each key to the ID of the SSTable where the latest version of the key is stored
so that only one filter check is needed by a point lookup.
To reduce the metadata overhead of SSTables, SlimDB uses a multi-level index structure as follows:
It first maps each prefix key into a list of pages that contain this prefix key
so that the key-value pairs can be retrieved efficiently given a prefix key.
It then stores the range of suffix keys for each page to efficiently
support point lookup queries based on both prefix and suffix keys.

Mathieu et al.~\cite{stack-merge2014} proposed two new merge policies optimized for
append-mostly workloads with a bounded number of components.
One problem of both leveling and tiering is that the number of levels depends on the total number of entries.
Thus, with an append-mostly workload, where the amount of data keeps increasing, the total number of levels will be unbounded in order to achieve the write cost described in \refsection{sec:lsm-cost}.
To address this, this work studied the theoretical lower bound of the write cost of an online merge policy for an append-mostly workload given at most $K$ components.
It further proposed two merge policies, MinLatency and Binomial, to achieve this lower bound.

The four improvements presented here each target a specialized workload.
It should be noted that their optimizations may be useless or even inapplicable for general purpose workloads.
For example, the LSM-trie~\cite{lsm-trie2015} only supports point lookups,
while SlimDB~\cite{slimdb2017} only supports a limited form of range queries by fetching all values for a prefix key.
The adoption of these optimizations should be chosen carefully based on the given workload.

\subsection{Auto-Tuning}
We now review some research efforts to develop auto-tuning techniques for the LSM-tree
to reduce the tuning burden for the end-user.
Some techniques perform co-tuning of all parameters to find an optimal design,
while others focus on some specific aspect such as merge policies, Bloom filters, or data placement.

\subsubsection{Parameter Tuning}
Lim et al.~\cite{lsm-model2016} presented an analytical model that incorporates the key distribution to improve the cost estimation of LSM-tree operations and further used this model to tune the parameters of LSM-trees.
The key insight is that the conventional worse-case analysis (\refsection{sec:lsm-cost}) fails to take the key distribution into consideration.
If a key is found to be deleted or updated during an early merge,
it will not participate in future merges and thus its overall write cost will be reduced.
The proposed model assumes a priori knowledge of the key distribution using a probability mass function $f_X(k)$
that measures the probability that a specific key $k$ is written by a write request.
Given $p$ total write requests, the number of unique keys is estimated using its expectation as 
$Unique(p) = N - \sum_{k\in K}{(1 - f_X(k))^p}$,
where $N$ is the total number of unique keys and $K$ is the total key space.
Based on this formula, the total write cost for $p$ writes can be computed by summing up the cost of all flushes and merges,
except that duplicates keys, if any, are excluded from future merges.
Finally, the cost model is used to find the optimal system parameters by minimizing the total write cost.

Monkey~\cite{monkey2017,monkey-tods} co-tunes the merge policy, size ratio, and memory allocation between memory components
and Bloom filters to find an optimal LSM-tree design for a given workload.
The first contribution of Monkey is to show that the usual Bloom filter memory allocation scheme,
which allocates the same number of bits per key for all Bloom filters, results in sub-optimal performance.
The intuition is that the $T$ components at the last level,
which contain most of the data, consume most of the Bloom filter memory but 
their Bloom filters can only save at most $T$ disk I/Os for a point lookup.
To minimize the overall false positive rates across all of the Bloom filters,
Monkey analytically shows that more bits should be allocated to the components at the lower levels so that
the Bloom filter false positive rates will be exponentially increasing.
Under this scheme, the I/O cost of zero-result point lookup queries will be dominated by the last level,
and the new I/O cost becomes $O(e^{-\frac{M}{N}})$ for leveling and $O(T\cdot e^{-\frac{M}{N}})$ for tiering.
Monkey then finds an optimal LSM-tree design by maximizing the overall throughput
using a cost model similar to the one in \refsection{sec:lsm-cost}
considering the workload's mix of the various operations.

\subsubsection{Tuning Merge Policies}
Dostoevsky~\cite{dostoevsky2018} shows that the existing merge policies, that is, tiering and leveling, are sub-optimal for certain workloads.
The intuition is that for leveling, the cost of zero-result point lookups,
long range queries, and space amplification are dominated by the largest level,
but the write cost derives equally from all of the levels.
To address this, Dostoevsky introduces a
lazy-leveling merge policy that performs tiering at the lower levels but leveling at the largest level.
Lazy-leveling has much better write cost than leveling, 
but has similar point lookup cost, long range query cost, and space amplification to leveling.
It only has a worse short range query cost than leveling since the number of components is increased.
Dostoevsky also proposes a hybrid policy that has at most $Z$ components
in the largest level and at most $K$ components at each of the smaller levels,
where $Z$ and $K$ are tunable.
It then finds an optimal LSM-tree design for a given workload using a similar method as Monkey~\cite{monkey2017}.
It is worth noting that the performance evaluation of Dostoevsky~\cite{dostoevsky2018} is very thorough;
it was performed against well-tuned LSM-trees to show
that Dostoevsky strictly dominates the existing LSM-tree designs under certain workloads.

Thonangi and Yang~\cite{partial-merge2017} formally studied the impact of partitioning on the write cost of LSM-trees.
This work first proposed a ChooseBest policy that always selects an SSTable with the fewest overlapping
SSTables at the next level to merge to bound the worst case merge cost.
Although the ChooseBest policy outperforms the unpartitioned merge policy in terms of the overall write cost,
there are certain periods when the unpartitioned merge policy has a lower write cost
since the current level becomes empty after a full merge,
which reduces the future merge cost.
To exploit this advantage of full merges, this work further proposed a mixed merge policy
that selectively performs full merges or partitioned merges based on the relative size between adjacent levels
and that dynamically learns these size thresholds to minimize the overall write cost for a given workload.

\subsubsection{Dynamic Bloom Filter Memory Allocation}
All of the existing LSM-tree implementations, even Monkey~\cite{monkey2017},
adopt a static scheme to manage Bloom filter memory allocation.
That is, once the Bloom filter is created for a component, its false positive rate remains unchanged.
Instead, ElasticBF~\cite{elastic-bf2018} dynamically adjusts the Bloom filter false positive rates based on the data
hotness and access frequency to optimize read performance.
Given a budget of $k$ Bloom filter bits per key, ElasticBF constructs multiple smaller Bloom filters with
$k_1$, $\cdots$, $k_n$ bits so that $k_1+\cdots+k_n = k$.
When all of these Bloom filters are used together, they provide the same false positive rate as the original monolithic Bloom filter.
ElasticBF then dynamically activates and deactivates these Bloom filters
based on the access frequency to minimize the total amount of extra I/O.
Their experiments reveal that ElasticBF is mainly effective when the overall Bloom filter memory is very limited,
such as only 4 bits per key on average.
In this case, the disk I/Os caused by the Bloom filter false positives will be dominant.
When memory is relatively large and can accommodate more bits per key, such as 10, the benefit of ElasticBF becomes limited
since the number of disk I/Os caused by false positives is much smaller than the number of actual disk I/Os to locate the keys.

\subsubsection{Optimizing Data Placement}
Mutant~\cite{mutant2018} optimizes the data placement of the LSM-tree on cloud storage.
Cloud vendors often provide a variety of storage options with different performance characteristics and monetary costs.
Given a monetary budget, it can be important to place SSTables on different storage devices properly to maximize system performance.
Mutant solves this problem by monitoring the access frequency of each SSTable
and finding a subset of SSTables to be placed in fast storage 
so that the total number of accesses to fast storage is maximized while the number of selected SSTables is bounded.
This optimization problem is equivalent to a 0/1 knapsack problem, which is N/P hard, and can be approximated using a greedy algorithm.

\subsubsection{Summary}
The techniques presented in this category aim at automatically tuning LSM-trees for given workloads.
Both Lim et al.~\cite{lsm-model2016} and Monkey~\cite{monkey2017,monkey-tods}
attempt to find optimal designs for LSM-trees to maximize system performance.
However, these two techniques are complimentary to each other.
Lim et al.~\cite{lsm-model2016} uses a novel analytical model to improve the cost estimation
but only focuses on tuning the maximum level sizes of the leveling merge policy.
In contrast, Monkey~\cite{monkey2017,monkey-tods}, as well as its follow-up work Dostoevsky~\cite{dostoevsky2018},
co-tune all parameters of LSM-trees to find an optimal design but only optimize for the worst-case I/O cost.
It would be useful to combine these two techniques together to enable more accurate performance tuning and prediction.

Dostoevsky~\cite{dostoevsky2018} extends the design space of LSM-trees with
a new merge policy by combining leveling and tiering.
This is very useful for certain workloads that require efficient writes, point lookups, and long range queries
with less emphasis on short range queries.
Thonangi and Yang~\cite{partial-merge2017} proposed to combine full merges with partitioned merges to achieve better write performance.
Other tuning techniques focus on some aspects of the LSM-tree implementation,
such as tuning Bloom filters by ElasticBF~\cite{elastic-bf2018} and optimizing data placement by Mutant~\cite{mutant2018}.

\subsection{Secondary Indexing}
So far, we have discussed LSM-tree improvements in a key-value store setting
that only contains a single LSM-tree.
Now we discuss LSM-based secondary indexing techniques to support efficient query processing, including index structures, index maintenance, statistics collection, and distributed indexing.

\begin{figure}
	\centering
	\includegraphics[width=0.9\linewidth]{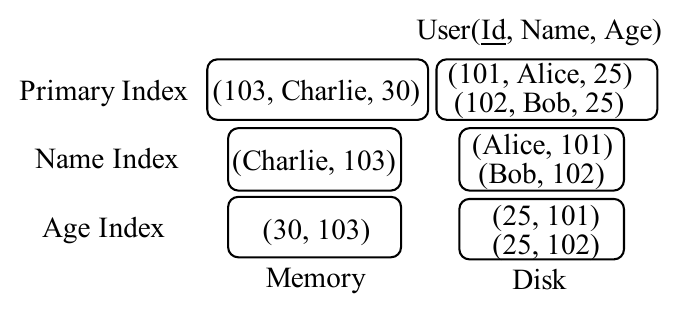}
	\vspace{-0.1in}
	\caption{Example LSM-based secondary indexes}
	\label{fig:secondary-index}
\end{figure}

Before we present these research efforts in detail,
we first discuss some basic concepts for LSM-based secondary indexing techniques.
In general, an LSM-based storage system will contain a primary index with multiple secondary indexes.
The primary index stores the record values indexed by their primary keys.
Each secondary index stores the corresponding primary keys for each secondary key using either a composite key approach
or a key list approach.
In the composite key approach, the index key of a secondary index is the composition of the secondary key and the primary key.
In the key list approach, a secondary index associates a list of primary keys with each secondary key.
Either way, to process a query using a secondary index, the secondary index is first searched to return a list of matching primary keys,
and those are then used to fetch the records from the primary index if needed.
An example of LSM-based secondary indexing is shown in \reffigure{fig:secondary-index}.
The example User dataset has three fields, namely Id, Name, and Age, where Id is the primary key.
The primary index stores full records indexed by Id, while the two secondary indexes store secondary keys, i.e., Name and Age, and their corresponding Ids.

\subsubsection{Index Structures}
The Log-Structured Inverted Index (LSII)~\cite{lsii2013} is an index structure designed for exact real-time keyword
search on microblogs. A query $q$ searches for the
top K microblogs with the highest scores, which are computed as the weighted sum of significance, freshness, and relevance.
To support efficient query processing, each keyword in a disk component stores three inverted lists of primary keys
in descending order of significance, freshness, and frequency, respectively.
Storing three inverted lists enables queries to be processed efficiently via the threshold algorithm~\cite{threshold2001}, which stops query evaluation once the upper bound of the scores of the unseen microbiologs is lower than the current top K answers.
However, only one inverted list is stored in the memory component since documents in the memory component often have high freshness and most of them will be accessed by queries.
Moreover, storing multiple inverted lists would significantly increase the memory component's write cost.

Kim et al.~\cite{lsm-spatial2017} conducted an experimental study of LSM-based spatial index structures for geo-tagged data,
including LSM-tree versions of the R-tree~\cite{rtree1984}, Dynamic Hilbert \btree (DHB-tree) \cite{space-filling2000},
Dynamic Hilbert Value \btree (DHVB-tree)~\cite{space-filling2000},
Static Hilbert \btree (SHB-tree)~\cite{sqlserver-spatial}, and Spatial Inverted File (SIF)~\cite{sif2010}.
An R-tree is a balanced search tree that stores multi-dimensional spatial data using their minimum bounding rectangles.
DHB-trees and DHVB-trees store spatial points directly into \btrees using space-filling curves.
SHB-trees and SIFs exploit a grid-based approach by statically decomposing a two-dimensional space into a multi-level grid hierarchy.
For each spatial object, the IDs of its overlapping cells are stored.
The difference between these two structures is that an SHB-tree stores the pairs of cell IDs and primary keys in a \btree,
while a SIF stores a list of primary keys for each cell ID in an inverted index.
The key conclusion of this study is that there is no clear winner among these index structures,
but the LSM-based R-tree performs reasonably well for both ingestion and query workloads without requiring too much tuning.
It also handles both point and non-point data well.
Moreover, for non-index-only queries,
the final primary key lookup step is generally dominant since it often requires a separate disk I/O for each primary key.
This further diminished the differences between these spatial indexing methods.

\begin{figure}
	\centering
	\includegraphics[width=1\linewidth]{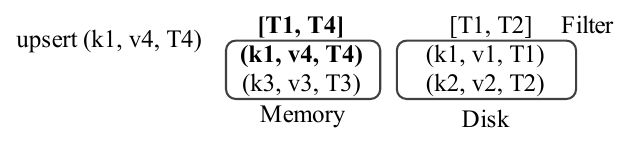}
	\vspace{-0.2in}
	\caption{Filter update example: the dataset contains three fields, a primary key (\emph{k}), a value (\emph{v}), and a creation time (\emph{T}) that is also the filter key}
	\label{fig:filter}
\end{figure}

Filters~\cite{asterixdb-filter2015} augment each component of the primary and secondary indexes with a filter to enable data pruning
based on a filter key during query processing.
A filter stores the minimum and maximum values of the chosen filter key for the entries in a component.
Thus, a component can be pruned by a query if the search condition is disjoint with the minimum and maximum values of its filter.
Though a filter can be built on arbitrary fields, it is really only effective for time-correlated fields since components are naturally partitioned based on time and are likely to have disjoint filter ranges.
Note that some special care is needed to maintain filters when a key is updated or deleted.
In this case, the filter of the memory component must be maintained based on both the old record and the new record
so that future queries will not miss new updates.
Consider the example in \reffigure{fig:filter}, which depicts a filtered primary LSM-tree.
After upserting the new record (k1, v4, T4), the filter of the memory component becomes [T1, T4]
so that future queries will properly see that the old record (k1, v1, T1) in the disk component has been deleted.
Otherwise, if the filter of the memory component were only maintained based on the new value T4, which would be [T3, T4],
a query with search condition T $\le$ T2 would erroneously prune the memory component
and thus actually see the deleted record (k1, v1, T1).

Qadar et al.~\cite{secondary2018} conducted an experimental study of LSM-based secondary indexing techniques
including filters and secondary indexes.
For filters, they evaluated component-level range filters and Bloom filters on secondary keys.
For secondary indexes, they evaluated two secondary indexing schemes based on composite keys and key lists.
Depending on how the secondary index is maintained, the key list scheme can be further classified as being either eager or lazy.
The eager key list scheme always reads the previous list to create a full new list with the new entry added
and inserts the new list into the memory component.
The lazy key list scheme simply maintains multiple partial lists at each component.
The experimental results suggest that the eager inverted list scheme incurs
a large overhead on data ingestion because of the point lookups and high write amplification.
When the query selectivity becomes larger, that is,
when the result set contains more entries, the performance difference between the lazy key list scheme 
and the composite key scheme diminishes, as the final point lookup step becomes dominant.
Finally, filters were found to be very effective with small storage overhead for time-correlated workloads.
However, the study did not consider cleaning up secondary indexes in the case of updates,
which means that secondary indexes could return obsolete primary keys.

\subsubsection{Index Maintenance}
A key challenge of maintaining LSM-based secondary indexes is handling updates.
For a primary LSM-tree, an update can blindly add the new entry (with the identical key) into the memory component so that the old entry is automatically deleted.
However, this mechanism does not work for a secondary index since a secondary key value can change during an update.
Extra work must be performed to clean up obsolete entries from secondary indexes during updates.

Diff-Index~\cite{diff-index2014} presents four index maintenance schemes for LSM-based secondary indexes,
namely sync-full, sync-insert, async-simple, and async-session.
During an update, two steps must be performed to update a secondary index,
namely inserting the new entry and cleaning up the old entry.
Inserting the new entry is very efficient for LSM-trees,
but cleaning up the old entry is generally expensive since it requires a point lookup to fetch the old record.
Sync-full performs these two steps synchronously during the ingestion time.
It optimizes for query performance since secondary indexes are always up-to-date,
but incurs a high overhead during data ingestion because of the point lookups.
Sync-insert only inserts new data into secondary indexes, while cleaning up obsolete entries lazily by queries.
Async-simple performs index maintenance asynchronously
but guarantees its eventual execution by appending updates into an asynchronous update queue.
Finally, async-session enhances async-simple with session consistency for applications by storing new updates temporarily
into a local cache on the client-side.

\zap{
\begin{table}
	\centering
	\caption{Summary of index update schemes supported by Diff-Index~\cite{diff-index2014}}
	\label{table:diff-index-update}
	\begin{tabular}{lm{2.5cm}m{2.6cm}}
		\toprule
		Scheme & Insert New Entry & Delete Old Entry \\
		\midrule
		sync-full & Sync & Sync \\
		sync-insert & Sync & Lazy (Query)\\
		async-simple & Lazy  & Lazy (Queued) \\
		async-session & Lazy & Lazy (Session Consistency) \\
		\bottomrule
	\end{tabular}
\end{table}
}

Deferred Lightweight Indexing (DELI)~\cite{deli2015}
enhances the sync-insert update scheme of Diff-Index~\cite{diff-index2014}
with a new method to cleanup secondary indexes by scanning the primary index components.
Specifically, when multiple records with identical keys are encountered when scanning primary index components,
the obsolete records are used to produce anti-matter entries to clean up the secondary indexes.
Note that this procedure can be naturally integrated with the merge process of the primary index to reduce the extra overhead.
Meanwhile, since secondary indexes are not always up-to-date,
queries must always validate search results by fetching records from the primary index.
Because of this, DELI cannot support index-only queries efficiently since point lookups must be performed for validation.

Luo and Carey~\cite{extended} presented several techniques for the efficient exploitation and maintenance of LSM-based auxiliary structures,
including secondary indexes and filters.
They first conducted an experimental study to evaluate the effectiveness of various point lookup optimizations,
including a newly proposed batched lookup algorithm that accesses components sequentially for a batch of keys,
stateful \btree search cursors, and blocked bloom filters~\cite{block-bloomfilter2009}.
They found that the batched lookup algorithm is the most effective
optimization for reducing random I/Os, while the other two are mainly effective for non-selective queries
at further reducing the in-memory search cost.
To maintain auxiliary structures efficiently, two strategies were further proposed.
The key insight is to maintain and exploit a primary key index, which only stores primary keys plus timestamps, to reduce disk I/Os.
A validation strategy was proposed to maintain secondary indexes lazily in the background, eliminating the synchronous point lookup overhead.
Queries must validate the primary keys returned by secondary indexes
either by fetching records directly from the primary index or by searching the primary key index to
ensure that the returned primary keys still have the latest timestamps.
Secondary indexes are cleaned up efficiently in the background using the primary key index to avoid accessing full records;
the basic idea for cleanup is to search the primary key index to validate whether
each secondary index entry still has the latest timestamp, as in query validation.
Compared to DELI~\cite{deli2015}, the validation strategy~\cite{extended} significantly reduces the I/O cost for cleaning up secondary indexes since only the primary key index is accessed.
A mutable-bitmap strategy was also introduced to efficiently maintain a primary index with filters.
It attaches a mutable bitmap to each disk component
so that old records can be directly marked as deleted, thereby avoiding the need to maintain filters based on old records.

\subsubsection{Statistics Collection}
Absalyamov et al.~\cite{lsm-cardinality2018} proposed a lightweight statistics collection framework for LSM-based systems.
The basic idea is to integrate the task of statistics collection into the flush and merge operations to minimize the statistics maintenance overhead.
During flush and merge operations, statistical synopses,
such as histograms and wavelets, are created on-the-fly and are sent back to the system catalog.
Due to the multi-component nature of LSM-trees, the system catalog stores multiple statistics for a dataset.
To reduce the overhead during query optimization, mergeable statistics, such as equi-width histograms, are merged beforehand.
For statistics that are not mergeable, multiple synopses are kept to improve the accuracy of cardinality estimation.

\begin{table*}
	\def\arraystretch{1.05}
	\centering
	\caption{Summary of trade-offs made by LSM-tree improvements ($\uparrow$ denotes increasing, $\downarrow$ denotes decreasing, $-$ denotes unaffected, and $\times$ denotes unsupported)}
	\begin{tabular}{lP{1cm}P{1.8cm}P{1.5cm}P{1.5cm}P{1.5cm}m{5cm}}
		\toprule
		Publication & Write & Point Lookup & Short Range & Long Range & Space & Remark \\
		\midrule
		WB-tree~\cite{wb-tree2013} &  $\uparrow\uparrow$ & $\downarrow$ & $\downarrow\downarrow$ & $\downarrow\downarrow$ & $\downarrow\downarrow$ & Tiering \\
		LWC-tree~\cite{lwc-tree-tos2017,lwc-tree2017} &  $\uparrow\uparrow$ & $\downarrow$ & $\downarrow\downarrow$ & $\downarrow\downarrow$ & $\downarrow\downarrow$ & Tiering \\
		PebblesDB~\cite{pebblesdb2017} &  $\uparrow\uparrow$ & $\downarrow$ & $\downarrow\downarrow$ & $\downarrow\downarrow$ & $\downarrow\downarrow$ & Tiering \\
		dCompaction~\cite{dcompaction2017} &  $\uparrow\uparrow$ & $\downarrow$ & $\downarrow\downarrow$ & $\downarrow\downarrow$ & $\downarrow\downarrow$ & Tiering \\
		Zhang et al.~\cite{groupedLSM2016} &  $\uparrow\uparrow$ & $\downarrow$ & $\downarrow\downarrow$ & $\downarrow\downarrow$ & $\downarrow\downarrow$ & Tiering \\
		SifrDB~\cite{sifrdb2018} &  $\uparrow\uparrow$ & $\downarrow$ & $\downarrow\downarrow$ & $\downarrow\downarrow$ & $\downarrow\downarrow$ & Tiering \\
		Skip-tree~\cite{skip-tree2017} &  $\uparrow$ & $\downarrow$ & $\downarrow$ & $\downarrow$ & $-$ & Mutable skip buffers \\
		TRIAD~\cite{triad2017} &$\uparrow$ &  $\downarrow$ & $\downarrow$ & $\downarrow$  & $-$ &
		Separate cold entries from hot entries; delay merges at level 0; use logs as flushed components \\
		VT-tree~\cite{vt-tree2013} & $\uparrow$ & $-$  & $\downarrow$ & $\downarrow$ & $\downarrow$ & Stitching merge \\
		MaSM~\cite{masm2011} & $\uparrow\uparrow$ & $\downarrow$ & $\downarrow\downarrow$ & $\downarrow$ & $\downarrow$
		& Lazy leveling \\
		WiscKey~\cite{wisckey2017} & $\uparrow\uparrow\uparrow$ & $\downarrow$ &$\downarrow\downarrow\downarrow$ & $\downarrow\downarrow\downarrow$ & $\downarrow\downarrow\downarrow$ & KV separation \\
		HashKV~\cite{hashkv2018} & $\uparrow\uparrow\uparrow$ & $\downarrow$ &$\downarrow\downarrow\downarrow$ & $\downarrow\downarrow\downarrow$ & $\downarrow\downarrow\downarrow$ &  KV separation \\
		Kreon~\cite{kreon2018}  & $\uparrow\uparrow\uparrow$ & $\downarrow$ &$\downarrow\downarrow\downarrow$ & $\downarrow\downarrow\downarrow$ & $\downarrow\downarrow\downarrow$ &  KV separation \\
		LSM-trie~\cite{lsm-trie2015} &$\uparrow\uparrow$  & $\uparrow$ & $\times $ &$\times$  & $\downarrow\downarrow$ & Tiering + hashing \\
		SlimDB~\cite{slimdb2017} & $\uparrow\uparrow$ & $\uparrow$  & $\downarrow\downarrow$/$\times$ & $\downarrow$/$\times$ & $\downarrow $&  Only support range queries for each key prefix group\\
		Lim et al.~\cite{lsm-model2016} & $\uparrow$ & $-$ & $-$ & $-$ & $-$ & Exploit data redundancy \\
		Monkey~\cite{monkey2017,monkey-tods} & $-$ & $\uparrow$ & $-$ & $-$ &$-$ &Better Bloom filter memory allocation \\
		Dostoevsky~\cite{dostoevsky2018} & $\uparrow\uparrow$ & $\downarrow$ & $\downarrow\downarrow$ & $\downarrow$ & $\downarrow$ & Lazy leveling \\
		\bottomrule
	\end{tabular}
	\label{table:summary-tradeoff}
\end{table*}

\subsubsection{Distributed Indexing}
Joseph et al.~\cite{local-global2017} described two basic implementations of
distributed secondary indexes on top of HBase~\cite{hbase},
namely global secondary indexes and local secondary indexes,
based on the two common approaches to indexing data in a parallel database.
A global secondary index is implemented as a separate table
that stores secondary keys plus their corresponding primary keys,
and it is maintained using co-processors provided by HBase (similar to database triggers).
This approach is easy to implement, but incurs a higher communication cost during data ingestion since a secondary index partition may be stored at a separate node from the primary index partition.
A local secondary index avoids the communication cost during data ingestion
by co-locating each secondary index partition together with the corresponding primary index partition.
However, the downside for HBase is that this approach has to be implemented from scratch.
Moreover, all partitions of a local secondary index must be searched, eve	n for highly selective queries,
since a local secondary index is partitioned by primary (not secondary) keys.

Zhu et al.~\cite{bulkload2017} introduced an efficient approach for loading global secondary indexes
using three steps:
First, the primary index at each partition is scanned and sorted to create a local secondary index.
Meanwhile, the statistics of the secondary key are collected to facilitate the next step.
Second, based on the collected statistics from the first stage, the index entries of the secondary index will be range-partitioned
and these partitions will be assigned to physical nodes.
Finally, based on the assigned secondary key range, 
each node fetches secondary keys and their primary keys from all other nodes,
which can be done efficiently by scanning the local secondary index built in the first stage.

Duan et al.~\cite{view2018} proposed a lazy maintenance approach for materialized views on distributed LSM-trees.
The basic idea is to append new updates into a delta list of the materialized view to reduce the overhead during data ingestion.
The changes in the delta list are then applied to the materialized view lazily, during query processing.

\subsubsection{Summary}
The techniques in this category all focus on improving LSM-trees in database settings with secondary indexes and other auxiliary structures.
Several LSM-based secondary indexing structures have been proposed,
including LSM-based inverted indexes~\cite{lsii2013}, spatial indexes~\cite{lsm-spatial2017}, and filters~\cite{asterixdb-filter2015}.
These structures would be helpful to optimize certain query workloads.
In terms of efficiently maintaining secondary indexes, a common approach is to defer the maintenance of secondary indexes so that expensive 
point lookups can be avoided during the ingestion time.
The proposed techniques mainly differ in how secondary indexes are cleaned up in the background, either by queries~\cite{diff-index2014}, scanning the primary index~\cite{deli2015}, or exploiting a primary key index~\cite{extended}.
Since the optimality of these methods may be workload dependent,
it would be useful as future work to design adaptive maintenance mechanisms to maximize performance.
Absalyamov et al.~\cite{lsm-cardinality2018} proposed a statistics collection framework,
which is a step towards cost-based query optimization on LSM-based systems.
Finally, several distributed indexing techniques~\cite{local-global2017,view2018,bulkload2017} have also been presented.
It should be noted that these techniques are not specific to LSM-trees, but we have included them here for completeness.

\subsection{Discussion of Overall Trade-offs}
Based on the RUM conjecture~\cite{rum2016}, no access method can be read-optimal, write-optimal, and space-optimal at the same time.
As we have seen in this survey, many LSM-tree improvements that optimize for certain workload or system aspects
will generally make trade-offs.
To conclude this section, we provide a qualitative analysis and summary of the trade-offs made
by those research efforts that seek to optimize various aspects of LSM-trees.
We will consider the leveling merge policy as the baseline for this discussion.

The performance trade-offs of the various LSM-tree improvements are summarized in~\reftable{table:summary-tradeoff}.
As one can see, most of these improvements try to improve the write performance of the leveling merge policy
since it has relatively high write amplification.
A common approach taken by existing improvements is to apply the tiering merge policy~\cite{wb-tree2013,sifrdb2018,dcompaction2017,pebblesdb2017,lsm-trie2015,lwc-tree-tos2017,lwc-tree2017,groupedLSM2016},
but this will negatively impact query performance and space utilization.
Moreover, tiering has a larger negative impact on range queries than point lookups since range queries cannot benefit from Bloom filters.

Other proposed improvements, such as the skip-tree~\cite{skip-tree2017}, TRIAD~\cite{triad2017}, and the VT-tree~\cite{vt-tree2013},
propose several new ideas to improve the write performance of LSM-trees.
However, in addition to extra overhead on query performance and space utilization,
these optimizations may bring non-trivial implementation complexity to real systems.
For example, the skip-tree introduces mutable buffers to store skipped keys, which contradicts the immutability of disk components.
TRIAD proposes to use transaction logs as disk components to eliminate flushes,
which is again highly non-trivial since transaction logs usually have
very different storage formats and operation interfaces from disk components.
Moreover, a common practice is to store transaction logs on a dedicated disk to minimize the negative impacts caused by log forces.
The stitching operation proposed by the VT-tree~\cite{vt-tree2013} can cause fragmentation and is incompatible with Bloom filters.

The LSM-trie~\cite{lsm-trie2015} and SlimDB~\cite{slimdb2017} give up some query capabilities to improve performance.
The LSM-trie exploits hashing to improve both read and write performance, but range queries cannot be supported.
SlimDB only supports a limited form of range queries based on a common prefix key.
These improvements would be desirable for certain workloads where complete range queries are not needed.

Separating keys from values~\cite{hashkv2018,wisckey2017,kreon2018} can drastically
improve the write performance of LSM-trees since only keys are merged.
However, a major problem is that range queries will be significantly impacted because values are not sorted anymore.
Even though this problem can be mitigated by exploiting the I/O parallelism of SSDs~\cite{hashkv2018,wisckey2017},
this still leads to lower disk efficiency especially when values are relatively small.
Moreover, storing values separately leads to lower space utilization since values are not garbage-collected during merges.
A separate garbage-collection process must be designed to reclaim disk space occupied by obsolete values.

Given that trade-offs are inevitable, it is valuable to explore the design space of
LSM-trees so that one can make better or optimal trade-offs.
For example, Lim et al.~\cite{lsm-model2016} exploits data redundancy to tune the maximum sizes for each level
to optimize write performance.
This has little or no impact on other performance metrics since the number of levels remains the same.
Another example is Monkey~\cite{monkey2017,monkey-tods}, which unifies the design space of LSM-trees 
in terms of merge policies, size ratios, and memory allocation between memory components and Bloom filters.
It further identifies a better memory allocation scheme for Bloom filters that improves point lookup performance without any negative impact on other metrics.
Finally, Dostoevsky~\cite{dostoevsky2018} extends the design space of LSM-trees with a new lazy leveling merge policy.
By performing tiering at lower levels while leveling at the largest level, lazy leveling achieves similar write throughput to tiering
but only has slightly worse point lookup performance, long range query performance, and space utilization than leveling.

\section{Representative LSM-based Systems}
\label{sec:lsm-systems}
Having discussed LSM-trees and their improvements in detail,
we now survey five representative LSM-based open-source NoSQL systems,
namely LevelDB~\cite{leveldb}, RocksDB~\cite{rocksdb}, Cassandra~\cite{cassandra},
HBase~\cite{hbase}, and AsterixDB~\cite{asterixdb2014}.
We will focus on their storage layers.

\subsection{LevelDB}
LevelDB~\cite{leveldb} is an LSM-based key-value store that was open-sourced by Google in 2011.
It supports a simple key-value interface including puts, gets, and scans.
LevelDB is not a full-fledged data management system, but rather an embedded storage engine
intended to power higher-level applications. The major contribution of LevelDB was
that it pioneered the design and implementation of the partitioned leveling merge policy,
which was described in \refsection{sec:lsm-basic-structure}.
This design has impacted many subsequent LSM-tree improvements
and implementations, as we have seen in this survey.
Since we have already described partitioned leveling in \refsection{sec:lsm-basic-structure}, we omit further discussions here.

\subsection{RocksDB}
RocksDB~\cite{rocksdb} was initially a fork of LevelDB created by Facebook in 2012.
Since then, RocksDB has added a large number of new features.
Due to its high performance and flexibility, RocksDB has successfully been used in various systems~\cite{rocksdb-space2017}
both inside and outside of Facebook.
According to Facebook, a major motivation of for their adoption of
LSM-based storage was its good space utilization~\cite{rocksdb-space2017}.
With the default size ratio of 10, RocksDB's leveling implementation has about 90\% percent of the total data at the largest level,
ensuring that at most 10\% of the total storage space can be wasted for storing obsolete entries.
As mentioned earlier, this outperforms traditional B-tree-based storage engines, where pages are typically 2/3 full on average due to fragmentation~\cite{btree-space-1978}.
Here we discuss various improvements made by RocksDB,
including its improvements to merge policies, merge operations, and new functionality.

RocksDB's LSM-tree implementation remains based on the partitioned leveling design, but with some improvements.
Since SSTables at level 0 are not partitioned, merging an SSTable from level 0 to level 1 generally causes
rewrites of all SSTables at level 1, 
which often makes level 0 the performance bottleneck.
To partially address this problem, RocksDB optionally merges SSTables at level 0 using the tiering merge policy.
This elastic design allows RocksDB to better absorb write bursts without degrading query performance too much.
RocksDB further supports a dynamic level size scheme to bound the space amplification.
The issue is that the ideal leveling space amplification $O(\frac{T+1}{T})$ is achieved only when the last level reaches the maximum size, which may not always happen in practice.
To address this, RocksDB dynamically adjusts the maximum capacities of all of the lower levels depending on the current size of the last level,
thereby ensuring that the space amplification is always $O(\frac{T+1}{T})$.
In addition to a round-robin policy to select the SSTables to be merged, which is used in LevelDB,
RocksDB supports two additional policies - namely cold-first and delete-first.
The cold-first policy selects cold SSTables to merge to optimize for skewed workloads.
It ensures that hot SSTables that are updated frequently will remain in the lower levels to reduce their total write cost.
The delete-first policy selects SSTables with a large number of anti-matter entries to quickly reclaim the disk space
occupied by the deleted entries.
Finally, RocksDB supports an API called the merge filter\footnote{It is called the \emph{compaction filter} in RocksDB since RocksDB prefers the term \emph{compaction} to \emph{merge}. We use the term \emph{merge} here
to minimize the potential for terminology confusion.} that allows users to provide custom logic to garbage-collect obsolete entries during merges efficiently.
During a merge, RocksDB invokes the user-provided merge filter with each key-value pair and 
only adds those key-value pairs that are not filtered to the resulting SSTables.

Besides the partitioned leveling merge policy, RocksDB supports other merge policies such as tiering and FIFO.
In RocksDB, as well as other systems, the actual tiering merge policy slightly differs from the one described in this paper (and elsewhere in the literature).
RocksDB's tiering merge policy is controlled by two parameters, namely, the number of components to merge (K) and the size ratio (T).
It works by examining components from oldest to newest, and for each component $C_{i}$, it checks whether the total size of the K-1 younger components $C_{i-1}$, $C_{i-2}$, ..., $C_{i-K}$ is larger than T times the size of $C_{i}$.
If so, the policy merges all of these components together; otherwise, it proceeds to check the next younger component.
RocksDB performs limited partitioning for its tiering merge policy, similar to the horizontal grouping design (\refsection{sec:lsm-optimizations}),
to bound the maximum size of SSTables.
The motivation is that the maximum page size is limited to 4GB, but the index page of a huge component (stored as a single SSTable)
could exceed this size limit.
However, during large merges the disk space may still be temporarily doubled since RocksDB treats each SSTable group as a whole
and only deletes old SSTables when the merge operation is fully completed.
In the FIFO merge policy, components are not merged at all,
but old components will be deleted based on a specified lifetime.

In LSM-based storage, merge operations typically consume a lot of CPU and disk resources
that can negatively impact query performance.
Moreover, the timing of merges is generally unpredictable, as it directly depends on the write rate.
To alleviate this issue, RocksDB supports rate limiting to control the disk write speed of merge operations based on the leaky bucket mechanism~\cite{leaky-bucket1986}.
The basic idea is to maintain a ``bucket''
that stores a number of tokens controlled by a token refill speed.
All flush and merge operations must request a certain number of tokens before performing each write.
Thus, the disk write speed of flush and merge operations will be bounded by the specified token refill speed.

Finally, RocksDB supports a new operation called read-modify-write.
In practice, many applications typically update existing values by reading them first.
To support this operation efficiently, RocksDB allows users to write delta records directly into memory,
thereby avoiding reading the original record.
Delta records are then combined with base records during query processing and merges based on the user-provided combination logic.
If applicable, RocksDB further combines multiple delta records together during
merges to improve subsequent query performance.

\subsection{HBase}
Apache HBase~\cite{hbase} is a distributed data storage system in the Hadoop ecosystem;
it was modeled after Google's Bigtable design~\cite{bigtable}.
HBase is based on a master-slave architecture.
It partitions (either hash or range) a dataset into a set of regions,
where each region is managed by an LSM-tree.
HBase supports dynamic region splitting and merging to elastically manage system resources based on the given workload.
Here we focus on the storage engine of HBase.

HBase's LSM-tree implementation is generally based on the basic tiering merge policy.
It supports some variations of the tiering merge policy as well,
such as the exploring merge policy and the date-tiered merge policy.
The exploring merge policy checks all mergeable component sequences and selects the one with the smallest write cost.
This merge policy is more robust than the basic tiering merge policy,
especially when components have irregular sizes due to loading and deletions.
Thus, it is used as the default merge policy in HBase.
The date-tiered merge policy is designed for managing time-series data.
It merges components based on their time ranges, instead of their sizes,
so that components will be time-range-partitioned.
This enables efficient processing of temporal queries.

Recently, HBase has introduced a new feature,
called stripping, to partition a large region to improve merge efficiency.
The idea is to partition the key space so that each partition, which contains a list of components, is merged independently.
This is similar to the design proposed by PE-files~\cite{pe-file2007},
but is different from the partitioned tiering merge policy described in \refsection{sec:lsm-basic-structure}.

HBase does not support secondary indexes natively.
However, a secondary index can be implemented as a separate table that stores secondary keys plus their primary keys 
using co-processors, as described in~\cite{local-global2017}.

\subsection{Cassandra}
Apache Cassandra~\cite{cassandra} is an open-source distributed data storage system modeled after both Amazon's Dynamo~\cite{dynamo2007} and Google's BigTable~\cite{bigtable}.
Cassandra relies on a decentralized architecture to eliminate the possibility of a single point of failure.
Each data partition in Cassandra is powered by an LSM-based storage engine.

Cassandra supports a similar set of merge policies to RocksDB and HBase,
including the (unpartitioned) tiering merge policy, the partitioned leveling merge policy, and the date-tiered merge policy.
Moreover, Cassandra supports local secondary indexes to facilitate query processing.
To avoid the high point lookup overhead, secondary indexes are maintained lazily, similar to DELI~\cite{deli2015}.
During an update, if the old record is found in the memory component, then it is used to clean up secondary indexes directly.
Otherwise, secondary indexes are cleaned up lazily when merging the primary index components.

\subsection{AsterixDB}
Apache AsterixDB~\cite{asterixdb2014} is an open-source Big Data Management System (BDMS) that aims to manage massive amounts of semi-structured (e.g., JSON) data efficiently.
Here we focus on the storage management aspect of AsterixDB~\cite{asterixdb-storage2014}.

AsterixDB uses a shared-nothing parallel database style architecture.
The records of each dataset are hash-partitioned based on their primary keys across multiple nodes.
Each partition of a dataset is managed by
an LSM-based storage engine, with a primary index, a primary key index, and multiple local secondary indexes.
AsterixDB uses a record-level transaction model to ensure that all of the indexes are kept consistent within each partition.
The primary index stores records indexed by primary keys, and the primary key index stores primary keys only.
The primary key index is built to support COUNT(*) style queries efficiently
as well as various index maintenance operations~\cite{extended} since it is much smaller than the primary index.

Secondary indexes use the composition of the secondary key and the primary key as their index keys.
AsterixDB supports LSM-based \btrees, R-trees, and inverted indexes using a generic LSM-ification framework
that can convert an in-place index into an LSM-based index.
For LSM-based R-trees, a linear order, such as a Hilbert curve for point data and a Z-order curve for non-point data,
is used to sort the entries in disk components,
while in the memory component, deleted keys are recorded in a separate \btree to avoid multi-path traversals during deletes.
AsterixDB also supports LSM-based inverted indexes to efficiently
process full-text queries and similarity queries~\cite{similarity2018}.
By default, each LSM index's components are merged independently using a tiering-like merge policy.
AsterixDB also supports a correlated merge policy that synchronizes the merges of all of a dataset's
indexes together to improve query performance with filters.
The basic idea of this policy is to delegate merge scheduling to the primary index.
When a sequence of primary index components are merged, all corresponding components from other indexes will be merged as well.

\section{Future Research Directions}
\label{sec:future-direction}
Categorizing and summarizing the existing LSM-tree improvements in the literature
reveals several interesting outages and opportunities for future work on LSM-based storage.
We now briefly discuss some future research directions suggested by the results of this survey.

\textbf{Thorough Performance Evaluation.}
As mentioned before, the tunability of LSM-trees has not been adequately considered in many of the research efforts to date.
Work on improvements has typically been evaluated against a default (untuned) configuration of LevelDB or RocksDB.
It is not clear how the improvements would compare against a well-tuned baseline LSM-tree for a given workload.
Moreover, many of the improvement proposals have primarily evaluated their impact on query performance,
with space utilization often being neglected.
This situation can be addressed in future research by more carefully considering the tunability of LSM-trees.

\textbf{Partitioned Tiering Structure.}
Tiering has been used by many LSM-tree improvements to reduce the write amplification of LSM-trees.
In \refsection{sec:lsm-basic-structure}, we have identified two possible partitioned tiering schemes,
namely horizontal grouping and vertical grouping, that
cover virtually all of the tiering-related LSM-tree improvements proposed recently.
However, the performance characteristics and trade-offs of these two schemes are not yet clear.
In general, vertical grouping permits more freedom when selecting SSTables to merge,
while horizontal grouping ensures that SSTables are fixed-size.
It would be useful as future work to systematically evaluate these two schemes and possibly design new schemes
that combine the advantages of both.

\textbf{Hybrid Merge Policy.}
Until recently, most LSM-tree improvements have assumed a homogeneous
merge policy of either leveling or tiering at all of the levels of an LSM-tree.
However, this has been shown to be sub-optimal for certain workloads~\cite{dostoevsky2018}.
A hybrid merge policy of leveling and tiering can provide much better write performance than leveling
with minimal impact on point lookups, long range queries, and space amplification.
As a future direction, it would be interesting to design and implement LSM-trees with hybrid merge policies
and revisit some of the key questions raised by this design choice.

\textbf{Minimizing Performance Variance.}
In practice, performance variance is as important a performance metric as absolute throughput.
Unfortunately, LSM-trees often exhibit large performance variances
because they decouple the in-memory writes from expensive background I/Os.
As we have seen in this survey,
bLSM~\cite{blsm2012} is the only attempt to minimize write stalls exhibited by LSM-trees.
However, bLSM itself still has several limitations.
It was designed just for the unpartitioned leveling merge policy,
and it only minimizes long write latencies caused by write stalls instead of the variance of the overall ingestion throughput.
As future work, it would be very useful to design mechanisms to minimize the performance variance of LSM-trees.

\textbf{Towards Database Storage Engines.}
Finally, most of the existing LSM-tree improvements have focused rather narrowly
on a key-value store setting involving a single LSM-tree.
As LSM-trees are gradually becoming widely used inside DBMS storage engines,
new query processing and data ingestion techniques
should be developed for this more general (multi-index) setting.
Possible examples include adaptive maintenance of auxiliary structures to facilitate query processing,
LSM-aware query optimization, and co-planning of LSM-tree maintenance tasks with query execution.

\section{Conclusion}
\label{sec:conclusion}
Recently, LSM-trees have become increasingly popular in modern NoSQL systems
due to advantages such as superior write performance, high space utilization,
immutability of on-disk data, and tunability.
These factors have enabled LSM-trees to be widely adopted and deployed
to serve a variety of workloads.

In this paper, we have surveyed the recent research efforts,
including efforts from both the database community and the systems community, to improve LSM-trees.
We presented a general taxonomy to classify existing LSM-tree improvements
based on the specific aspects that they aim to optimize,
and we discussed the improvements in detail based on the proposed taxonomy.
We also reviewed several representative LSM-based open-source NoSQL systems,
and we identified some interesting future research directions.
We hope that this survey will serve as a useful guide to the state-of-the-art in LSM-based storage techniques
for researchers, practitioners, and users.

\begin{acknowledgements}
We would like to thank Mark Callaghan, Manos Athanassoulis,
and the anonymous reviewers for their valuable comments and feedback.
This work was supported by NSF awards CNS-1305430, IIS-1447720, and IIS-1838248 along with industrial support from Amazon, Google, and Microsoft and support from The Donald Bren Foundation (via a Bren Chair) of UC Irvine.
\end{acknowledgements}

\zap{
\section{Disclosure of potential conflicts of interest}
Funding: This study was funded by NSF awards CNS-1305430, IIS-1447720, and IIS-1838248 along with industrial support from Amazon, Google, and Microsoft and support from The Donald Bren Foundation (via a Bren Chair).

Conflict of Interest: Author Chen Luo has employer-based conflicts with University of California, Irvine. Author Michael J. Carey has employer-based conflicts with University of California, Irvine and Couchbase Inc.
}

\bibliographystyle{spmpsci}      
\bibliography{lsm}   

\end{document}